\documentclass[default,iicol]{sn-jnl}

\usepackage[utf8]{inputenc}
\usepackage{placeins}
\usepackage{upgreek}
\usepackage[version=4]{mhchem}
\usepackage{titlesec}
\usepackage[switch]{lineno}
\usepackage{tocloft}
\usepackage{cuted}
\usepackage[labelformat=empty,font=small]{caption}

\jyear{2026}

\theoremstyle{thmstyleone}

\theoremstyle{thmstyletwo}

\theoremstyle{thmstylethree}

\titleformat{\section}
  {\normalfont\large\bfseries\raggedright}
  {S.I.~\thesection}
  {1em}
  {}
\titlespacing*{\section}{0pt}{*3}{*1}

\usepackage{graphics}

\DeclareUnicodeCharacter{2212}{-}

\begin{document}

\title[Article Title]{Demonstration of on-chip all-optical switching of magnetization in integrated photonics}

\author[1,3]{\fnm{Pingzhi} \sur{Li}}

\author*[1]{\fnm{Gijs W.A.} \sur{Simons}}\email{g.w.a.simons@tue.nl}
\author[1]{\fnm{Tianyu} \sur{Zhang}}

\author[2]{\fnm{Philip P.J.} \sur{Schrinner}}
\author[2]{\fnm{Sohrab} \sur{Kamyar}}
\author[2]{\fnm{Ronald} \sur{Dekker}}

\author[1]{\fnm{Diana C.} \sur{Leitao}}
\author[1]{\fnm{Reinoud} \sur{Lavrijsen}}
\author[3]{\fnm{Yuqing} \sur{Jiao}}
\author[1]{\fnm{Bert} \sur{Koopmans}}

\affil[1]{\orgdiv{Department of Applied Physics}, \orgname{Eindhoven University of Technology}, \orgaddress{\city{Eindhoven}, \postcode{5612 AZ}, \country{Netherlands}}}
\affil[2]{\orgname{LioniX International BV}, \orgaddress{\city{Enschede}, \postcode{7500 AL}, \country{Netherlands}}}
\affil[3]{\orgdiv{Department of Electrical Engineering}, \orgname{Eindhoven University of Technology}, \orgaddress{\city{Eindhoven}, \postcode{5612 AZ}, \country{Netherlands}}}

\abstract{
    \textbf{\noindent Ultrafast all-optical magnetization switching (AOS) holds great promise for next-generation spintronic memory and hybrid spintronic-photonic systems. However, most implementations to date rely on bulky free-space optical setups, limiting scalability and practical integration. As a critical step toward integrated applications, we demonstrate single-pulse AOS within a silicon nitride (\ce{Si3N4}) photonic integrated circuit. Using trains of femtosecond laser pulses guided through on-chip waveguides, we achieve deterministic toggle switching in a sub-micron out-of-plane \ce{Co}/\ce{Gd} Hall cross patterned directly atop the photonic waveguide. Electrical readout via the anomalous Hall effect reveals a switching contrast of up to $90\%$ for 500~nm-wide devices. In larger Hall crosses, the contrast decreases and switching becomes stochastic, consistent with spatially non-uniform optical absorption as confirmed by finite-element simulations. This behavior is hypothetically attributed to domain wall relaxation and thermally assisted (de)pinning processes within partially switched regions. Our results highlight the critical role of device scaling in achieving robust on-chip AOS and establish a foundation for ultrafast, energy-efficient, and fully integrated spintronic-photonic platforms.
    }
}


\maketitle
\newpage

\noindent As global data consumption continues to accelerate, there is growing interest in developing spintronic devices that offer non-volatile memory and logic functionalities with high endurance and speed \cite{dieny_opportunities_2020, nguyen_recent_2024, mishra_voltage-controlled_2024, parkin_memory_2015-1}. In spintronic memory, information is written and read by controlling and sensing the magnetic state of nanoscale ferromagnets. Binary data is encoded by switching the magnetization between two stable orientations. However, conventional spintronic driven switching mechanisms \cite{ralph_spin_2008, choe_recent_2023, nguyen_recent_2024}
are fundamentally constrained by spin precession, which eventually impose limits on its energy efficiency and switching speed.

\begin{figure*}[t]
    \centering
    \includegraphics[width=1\linewidth]{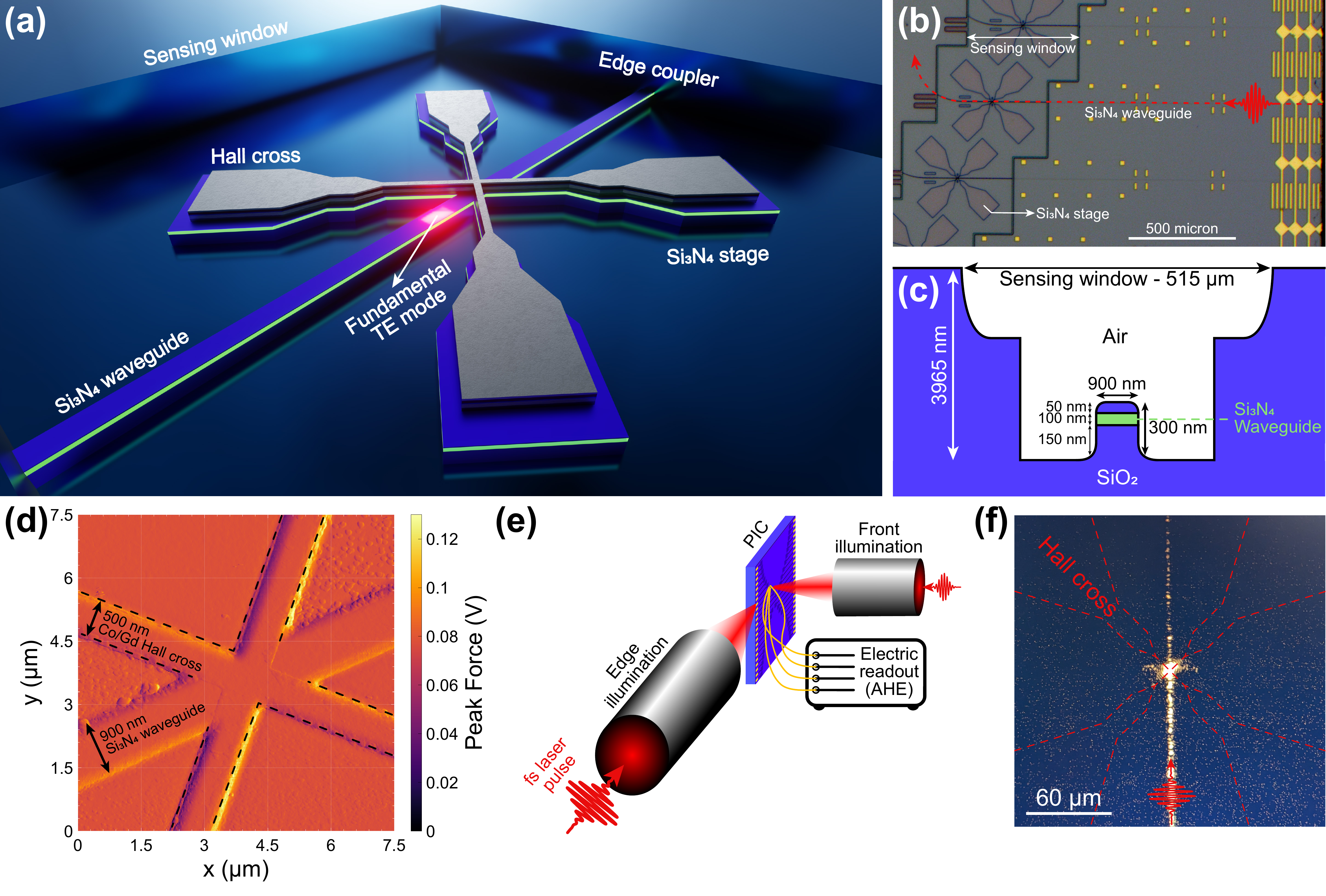}
    \caption{\textbf{Fig. 1 $\mid$ Schematic and microscopic characterization of the integrated spintronic–photonic device.} \textbf{(a)} Conceptual illustration of the device architecture, showing a patterned Hall cross structure integrated atop a \ce{Si3N4} waveguide. Extended \ce{Si3N4} stages are included to ensure a planar device. \textbf{(b,c)} Fabrication details of the sensing window: \textbf{(b)} Optical microscope image highlighting the \ce{Si3N4} waveguide, the surrounding stages, and the sensing window. \textbf{(c)} Cross-sectional schematic of the sensing window, illustrating the selectively etched \ce{SiO2} cladding. \textbf{(d)} Atomic force microscopy (AFM) image showing topographical details of the final device structure, with a Hall cross arm width of 500~nm aligned to the integrated \ce{Si3N4} waveguide and stage. \textbf{(e)} Conceptual illustration of the experimental setup for on-chip all-optical switching (AOS) measurements. A beam of femtosecond laser pulses is focused onto an edge coupler (\textit{edge illumination}) to inject light into the \ce{Si3N4} waveguide, or orthogonally onto the Hall cross surface (\textit{front illumination}). Optical switching of the Hall cross located on top of the waveguide is electrically detected via the anomalous Hall effect (AHE). See Methods for further details. \textbf{(f)} Top view of the device (dashed lines) imaged by a CCD camera, in the presence of high repetition rate edge illumination and absence of back light, which visualizes light leakage radiating from the waveguide, with intense scattering observed at the Hall cross section.}
    \label{fig:LayoutOnchipAOS}
\end{figure*}

Since its initial discovery \cite{beaurepaire_ultrafast_1996}, the field of ultrafast magnetism has not only posed exciting challenges for fundamental physics, but also revealed powerful mechanisms for ultrafast spintronic applications \cite{walowski_perspective_2016, kimel_writing_2019, bull_spintronic_2021, wang_picosecond_2022, Salomoni_field_2023}. Examples range from magnetization reversal driven by picosecond or sub-picosecond optical pulses to ultrafast switching triggered by optically generated electrical pulses {\cite{yang_ultrafast_2017}} or spin–orbit torques {\cite{polley_picosecond_2023}}. In particular, single-pulse all-optical switching (AOS) of magnetization \cite{stanciu_all-optical_2007, ostler_ultrafast_2012} has emerged as a promising route toward the fastest and one of the most energy-efficient magnetization reversal \cite{yang_ultrafast_2017, kimel_writing_2019}. Optical approaches also allow for direct magnetization readout, such as through the magneto-optical Kerr effect (MOKE) \cite{zhang_paradigm_2009, demirer_integrated_2022}, eliminating the need for energy-intensive electronic conversion and opening the door to fully spintronic-photonic platforms.

Recent advances in AOS, although demonstrated only with free-space optical setups, have significantly deepened our understanding of the fundamental mechanisms \cite{koopmans_explaining_2010, radu_transient_2011} and enabled optimization of key performance metrics, including repetition rate \cite{alebrand_interplay_2012, parlak_optically_2018, van_hees_toward_2022}, energy efficiency \cite{wang_enhanced_2020, li_ultra-low_2021}, and pulse duration \cite{wei_all-optical_2021, peng_in-plane_2023, verges_extending_2024, li_picosecond_2025}. These efforts have laid the groundwork for translating AOS into practical spintronic devices.

Integration of optically switchable magnetic layers into functional spintronic architectures has already been demonstrated in several works, including magnetoresistive random-access memory (MRAM) \cite{wang_picosecond_2022, Salomoni_field_2023}, where magnetic tunnel junctions (MTJs) offer high resistance readout contrast. AOS-based MRAM has shown switching speed improvements by up to three orders of magnitude compared to conventional spintronic writing approaches \cite{wang_picosecond_2022, nguyen_recent_2024}. Furthermore, AOS has been proposed as an extension to the racetrack memory framework \cite{lalieu_integrating_2019}, in which data is stored and transported after optical writing via current-driven magnetic domain walls along a magnetic strip \cite{parkin_memory_2015-1, li_ultrafast_2023, pezeshki_integrated_2023}. The recent demonstration of AOS coexisting with efficient domain wall motion in \ce{Co}/\ce{Gd}-based racetracks \cite{li_ultrafast_2023} marks a notable step forward. However, in all of these demonstrations, optical excitations were still delivered via free-space optics. These studies acknowledged, yet left unresolved, the central challenge of achieving efficient photon delivery in a compact and integrated form, a task that is inherently non-trivial in both design and implementation.

Integrated photonics offers a compelling solution to these limitations by providing compact, scalable, and energy-efficient on-chip optical routing and signal manipulation. Originally developed for datacom and telecom applications \cite{siew_review_2021, wang_scaling_2024}, integrated photonics relies on lithographically defined photonic waveguide networks fabricated on a common platform \cite{miller_integrated_1969}. This technology provides an ideal foundation for embedding AOS monolithically onto a chip, thereby unlocking its full potential for large-scale integration \cite{becker_out_2019, kimel_writing_2019}.

In this context, integrated photonic memories -- such as optical data buffers -- are actively being explored \cite{alexoudi_optical_2020, youngblood_integrated_2023}, and spintronics presents an attractive path toward fast and energy-efficient memory solutions. As such, hybrid spintronic-photonic integration is drawing increasing interest from both research communities.

Despite this progress, significant challenges remain for the monolithic integration of spintronic systems with ultrafast photonic architectures. Maintaining the integrity of femtosecond (fs) pulses during propagation requires careful control of dispersion and nonlinear optical losses \cite{pezeshki_integrated_2023, li_picosecond_2025}. Furthermore, ensuring efficient and balanced spatially resolved energy transfer from photonic to spintronic domains demands precise engineering of the material interfaces and electromagnetic field confinement \cite{becker_out_2019, li_ultra-low_2021, pezeshki_integrated_2023, pezeshki_integrated_2024}. Despite ongoing efforts, a critical milestone has yet to be achieved: the experimental demonstration of AOS using guided, confined optical pulses within a fully integrated photonic circuit.

In this work, we report the experimental demonstration of single-pulse AOS driven by fs laser pulses guided within a photonic integrated circuit (PIC), as illustrated in Fig.~\textcolor{blue}{1}. The platform is based on silicon nitride (\ce{Si3N4}) waveguides fabricated using the LioniX TriPleX technology \cite{roeloffzen_low-loss_2018}, which benefits from the wide and indirect bandgap of \ce{Si3N4} to suppress nonlinear losses \cite{kaur_hybrid_2021}, a key consideration when using fs pulses in nanophotonic structures \cite{pezeshki_integrated_2023}. Trains of fs laser pulses are coupled into the \ce{Si3N4} waveguides via on-chip edge couplers, enabling robust all-optical toggle switching of magnetization in an out-of-plane ferrimagnetic \ce{Co}/\ce{Gd}-heterostructure patterned into a Hall cross geometry. This design allows stable electrical readout of the magnetic state via the anomalous Hall effect (AHE). Moreover, by examining Hall crosses of varying dimensions, we observed a size-dependent transition from deterministic to stochastic switching behavior. Finite element simulations attribute this trend to non-uniform light absorption across the device, in combination with a hypothesis including domain wall relaxation and (de)pinning phenomena.

\section*{Device design}\label{Sec: DeviceDesign}
For the spintronic material stack, we used \ce{Ta}(2)/\ce{Pt}(2)/\ce{Co}(1)/\ce{Gd}(3)/\ce{TaN}(5)/\ce{Pt}(2), where the numbers in parentheses denote the layer thicknesses in nanometers. This stack is hereafter referred to as \ce{Co}/\ce{Gd}. This choice is motivated by its relevance in state-of-the-art spintronics, combining well-defined perpendicular magnetic anisotropy, strong spin–orbit effects, and ferrimagnetic ordering. These properties are essential for achieving deterministic all-optical switching (AOS) \cite{lalieu_deterministic_2017} and high current-driven domain wall velocities \cite{li_ultrafast_2023}. The \ce{TaN} layer is included to minimize intermixing between the \ce{Ta} and \ce{Gd} layers, which can degrade the stability of its magneto-static properties and AOS performance \cite{kools_aging_2023}.

The stack was lithographically patterned into a Hall cross device to enable electrical readout of magnetization in the cross region. The Hall cross was aligned and overlaid atop an integrated \ce{Si3N4} waveguide, as conceptually illustrated in Fig.~\textcolor{blue}{1a}.

To ensure sufficient light–matter interaction between the guided optical mode and the spintronic Hall cross, the few-microns-thick protective \ce{SiO2} cladding layer was selectively etched from the top surface and sidewalls in regions where Hall crosses were fabricated. These etched regions, commonly referred to as sensing windows \cite{roeloffzen_low-loss_2018}, are highlighted in Fig.~\textcolor{blue}{1b}. Approximately 50~nm of residual dielectric cladding remains atop the \ce{Si3N4} waveguide in these regions. Atomic force microscopy (AFM) scans were performed across multiple sensing windows to obtain detailed topographical information, summarized schematically in Fig.~\textcolor{blue}{1c}.

To maintain smooth electrical conduction within the Hall cross, we introduced \ce{Si3N4} stages extending laterally from the waveguide with the same thickness, serving as a structural template for the device. The final integrated Hall cross, positioned atop both the \ce{Si3N4} waveguide and stage, is shown in the AFM image in Fig.~\textcolor{blue}{1d}. Further details on the fabrication process and magnetostatic considerations are provided in the Methods section and Ref.~\cite{kools_aging_2023}, respectively.

A conceptual illustration of the experimental setup is shown in Fig.~\textcolor{blue}{1e}. Linearly polarized femtosecond (fs) laser pulses at a wavelength of 720~nm are focused onto embedded edge couplers \cite{almeida_nanotaper_2003, roeloffzen_low-loss_2018, leinse_triplex_2013} using an external objective in free space. The polarization is adjusted to maximize coupling efficiency to the fundamental transverse electric (TE) mode of the \ce{Si3N4} waveguide. Alternatively, the fs-laser pulses can be focused directly onto the Hall cross to enable \textit{front illumination}, allowing for comparative studies with \textit{edge illumination} via the coupler.

During illumination, an alternating current is applied to one arm of the Hall cross, and the voltage induced by the anomalous Hall effect (AHE) is electrically detected using a lock-in amplifier (see Methods for details). In addition to monitoring optical switching, the AHE signal serves as feedback for aligning the free-space laser pulses onto the edge coupler.

Sufficient coupling of light into the \ce{Si3N4} waveguide is visualized in Fig.~\textcolor{blue}{1f}, where guided light is observed leaking from the waveguide due to material inhomogeneities. Notably, a significant portion of the light propagating from bottom (upstream) to top (downstream) is either scattered or absorbed at the Hall cross section (outlined in red), as indicated by a pronounced drop in the intensity of the leaking light after this interaction. It is this light–matter interaction that triggers the magnetization switching.

\section*{On-chip all-optical switching}\label{Sec: OnChipAOS}
\begin{figure*}
    \centering
    \hspace*{-0.5cm}
    \includegraphics[width=1\linewidth]{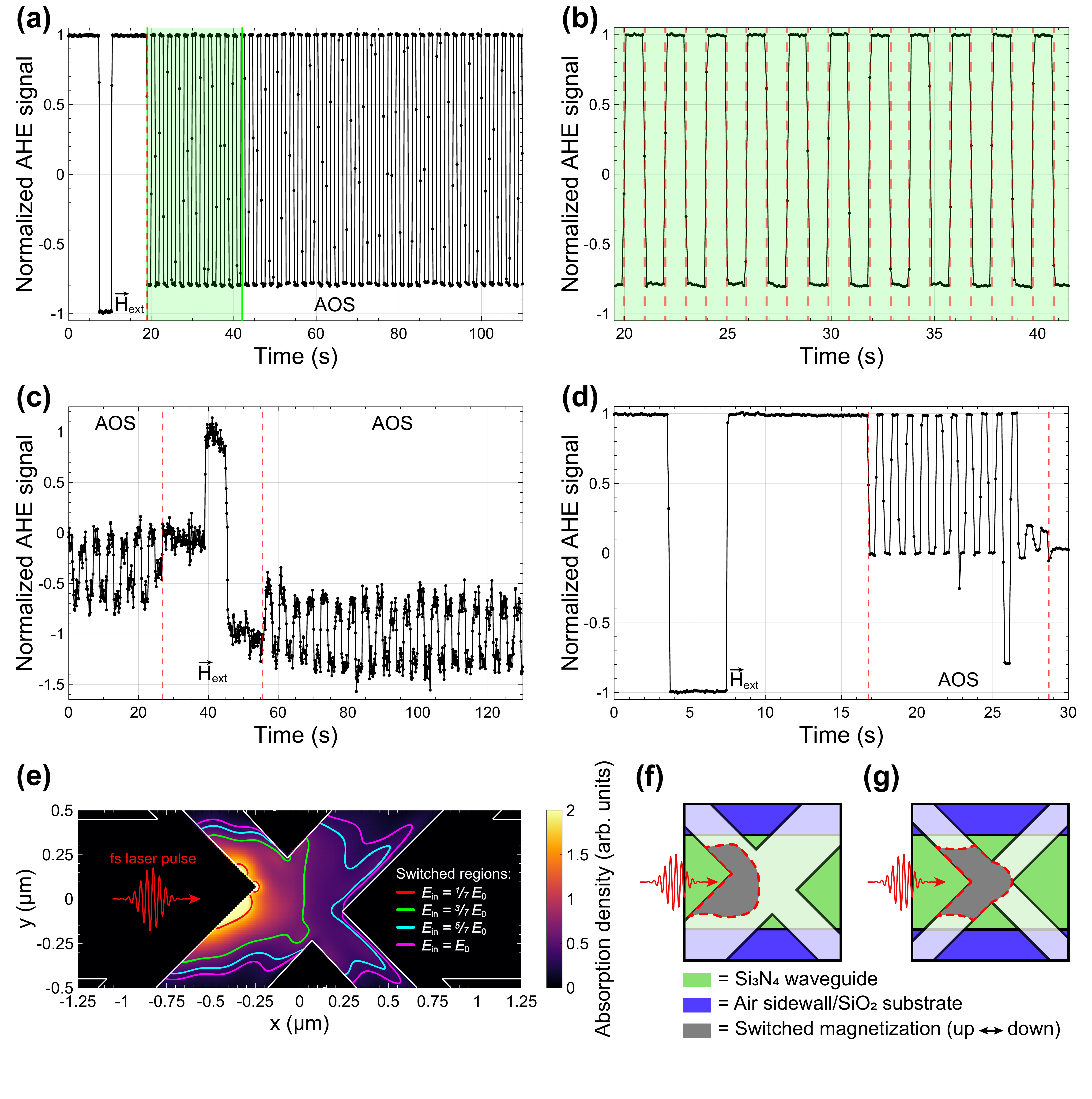}
    \caption{\textbf{Fig. 2 $\mid$ Analyses of experimental and simulated data for on-chip all-optical magnetization switching.} \textbf{(a)} Time trace of the normalized anomalous Hall effect (AHE) signal from an on-chip 500~nm Hall cross device. The signal is normalized to the signal swing of a full magnetization switch induced by an external out-of-plane magnetic field, $\vec{H}_{\text{ext}}$. After the dashed line, $\vec{H}_{\text{ext}}$ is removed and the device is illuminated with a train of femtosecond (fs) laser pulses at a repetition rate of 1~Hz. \textbf{(b)} Magnified view of the highlighted region in (a), showing periodic toggle switching of a specific Hall cross region induced by fs-laser pulses at 1~Hz. \textbf{(c)} Time trace of the normalized AHE signal showing partial magnetization reversal in a 1~$\mathrm{\upmu}$m Hall cross device under similar laser excitation conditions. \textbf{(d)} Normalized AHE signal measured near the switching threshold energy ($0.85 E_{0}$) in the 500~nm Hall cross. At this threshold, fs-laser pulse illumination at 1~Hz causes intermittent switching, zero crossings, and compensated magnetic states. Minor first-order drift corrections were applied to the signals in (a) and (c); further details are provided in S.I.~\textcolor{blue}{A}. \textbf{(e)} Finite-difference time-domain (FDTD) simulation of the spatially imbalanced absorption profile in the \ce{Co} layer, illustrating the non-uniform energy distribution that contributes to the partial switching and thermal damage observed in (a–d). The absorption density is normalized to the optical power in the waveguide away from the Hall cross. Switchable regions, calculated using a simplified microscopic three-temperature model, are indicated for different optical input energies, $E_{\text{in}}$. \textbf{(f,g)} Conceptual schematics of probable magnetic domain wall formation and pinning at the device edges: \textbf{(f)}, for the 1~$\mathrm{\upmu}$m Hall cross, corresponding to (c); \textbf{(g)}, for the 500~nm Hall cross, corresponding to (a), (b), and (d), consistent with the observed stochastic switching behavior.}
    \label{fig:AOSResults}
\end{figure*}
We first focus on the proof-of-concept demonstration of on-chip all-optical switching, measured using a symmetric Hall cross device with an arm width of 500~nm. After precisely aligning a focused train of fs laser pulses at a 1~Hz repetition rate onto the edge coupler, and increasing the pulse energy well above the magnetization switching threshold, we observed a clear toggling of the anomalous Hall effect (AHE) signal, synchronous with the incoming pulses (Fig.~\textcolor{blue}{2a}). This behavior evidences robust, all-optical toggle switching of the magnetization within the Hall cross with each successive laser pulse.

The normalized Hall signal toggled with an amplitude reaching approximately $90\%$ of the signal swing achieved under an externally applied out-of-plane magnetic field $\vec{H}_{\text{ext}}$, indicating nearly complete magnetization reversal of the active Hall cross region by the fs laser pulses. This asymmetry between the positive and negative AHE levels arises because the device is initially prepared in a fully saturated magnetic ``up''-state (along the {$+z$}-direction) by the external field, whereas all-optical switching reverses the Hall-sensitive region incompletely, resulting in an AHE signal that consistently toggles between a fully saturated ({$+1$}) and a partially reversed state ({$-0.8$}). A magnified view of a selected time interval (Fig.~\textcolor{blue}{2b}) reveals this fully deterministic toggling between two well-defined states. This periodic switching is consistent with established single-pulse all-optical switching behavior \cite{stanciu_all-optical_2007, radu_transient_2011, ostler_ultrafast_2012, lalieu_deterministic_2017}, and demonstrates the feasibility of fully integrated, reversible magnetization control via ultrafast optical pulses.

Repeating the same procedure on a device with an $1100 \times 700$ nm$^{2}$ asymmetric arm width (termed the 1~$\mathrm{\upmu}$m device, with further details provided in the Methods section) yielded a similar toggle-switching pattern (Fig.~\textcolor{blue}{2c}), but with a substantially reduced signal amplitude -- from $90\%$ in the 500~nm device to no more than $30\%$ in the 1~$\mathrm{\upmu}$m case. Despite this reduced contrast, the reproducibility of the switching remains evident for the specific protocol employed. After 25 seconds of repetitive laser pulses, the laser is switched off, and the Hall signal stabilizes. Applying a positive external magnetic field resets the device into a fully “up” magnetized state, followed by complete reversal under a negative field. Upon reactivating the laser pulse train, the original $30\%$ switching signal reappears, confirming the repeatability of the optical control.

To understand the reduced switching contrast in the larger device, we performed Ansys Lumerical finite-difference time-domain (FDTD) 3D electromagnetic simulations of the absorption profile for guided laser light interacting with the magnetic cladding (see Methods). The results, shown in Fig.~\textcolor{blue}{2e}, reveal a highly inhomogeneous absorption density profile. Contour lines indicate the calculated switching boundary as a function of optical input power $E_{\text{in}}$, where $E_{0}$ is the energy required to induce full switching of the entire central region of the Hall cross, based on a simplified microscopic three-temperature model (M3TM) \cite{koopmans_explaining_2010, beens_comparing_2019} (see supplementary information (S.I.)~\textcolor{blue}{B} for more detailed discussions). The inhomogeneity arises partly from attenuation of the propagating light, but is further exacerbated by local plasmonic field enhancements at the sharp edges of the Hall cross (left side) as well as interference effects. As a result, achieving the local threshold fluence required for AOS on the downstream side is challenging without risking thermal damage at the upstream edge.

These simulations link the reduced switching contrast in the 1~$\mathrm{\upmu}$m device to partial magnetization reversal, as schematically exemplified in Fig.~\textcolor{blue}{2f}, in contrast to the near-complete switching observed in the 500~nm device (Fig.~\textcolor{blue}{2g}). The gray shaded areas in the schematics denote the switched regions, with the red dashed line in Fig~\textcolor{blue}{2g} marking the nearly complete toggle region in the smaller device.

Larger devices are more susceptible to this inhomogeneous absorption profile. Achieving uniform switching across the full Hall cross requires significantly higher pulse energies, which increases the risk of thermal damage. This leads to local heating that degrades the AHE signal amplitude, resulting in noisier signals and drift effects, as observed in Fig.~\textcolor{blue}{2c}.

Further insights emerge when reducing the pulse energy in the smaller 500~nm device. In this case, the AHE signal evolves from deterministic toggle switching with a full $90\%$ amplitude to an increasingly stochastic pattern with intermediate states, as represented in Fig.~\textcolor{blue}{2d} (Additional datasets for varied input energies are provided in S.I.~\textcolor{blue}{C}). The first few pulses induce partial magnetization switching of approximately $50\%$, corresponding to a near-compensated magnetic state between “up” and “down”. Subsequent pulses reinitialize the magnetic state, ruling out complete demagnetization. After several pulses, the magnetization crosses zero -- similar to the behavior at higher pulse energies -- before settling into an intermediate state with nearly compensated domains.

These observations indicate that the transition from deterministic to stochastic switching is intimately linked to both device geometry and the complex spatial profile of optical energy deposition. In the following section, we explore the underlying mechanisms driving this stochastic behavior in more detail, focusing on the role of domain wall formation, pinning, and the interplay between local energy absorption and magnetic dynamics at timescales well above those of AOS.

\section*{Magnetic domain relaxation}\label{Sec: MagneticDomainRelaxation}
To investigate the origins of the stochastic switching behavior in more detail, we conducted a series of experiments under conditions that produce a more uniform optical absorption profile. This was achieved by employing \textit{front illumination} (see Fig.~\textcolor{blue}{1e}) at normal incidence on the 1~$\mathrm{\upmu}$m Hall cross device ($1100 \times 700$ nm$^{2}$ asymmetric Hall cross). Operating the fs-laser at a repetition rate of 1~Hz, we confirmed that a specific pulse energy, $E_{0}$, induces deterministic toggle switching. The corresponding AHE signal consistently reaches $100\%$ amplitude, indicating complete magnetization reversal of the active Hall cross region (Fig.~\textcolor{blue}{3a}).

When the pulse energy is reduced to $0.81 E_{0}$, stochastic switching behavior begins to emerge. While several successive pulses still produce full toggle switching, the system occasionally transitions to an intermediate state -- approximately $65\%$ of the maximum AHE signal in this case. During certain time intervals, the toggling continues between two complementary intermediate states before spontaneously returning to full amplitude toggling. As the pulse energy is reduced further, e.g. to $0.67 E_{\text{0}}$, the switching becomes increasingly stochastic. This stochasticity is reflected in the occasional failure of the magnetic domain to switch under incident laser pulses (indicated by dashed red lines) and in the emergence of more varied magnetic states.

\begin{figure}[t]
    \centering
    \hspace*{-0.5cm} 
    \includegraphics[width=1\linewidth]{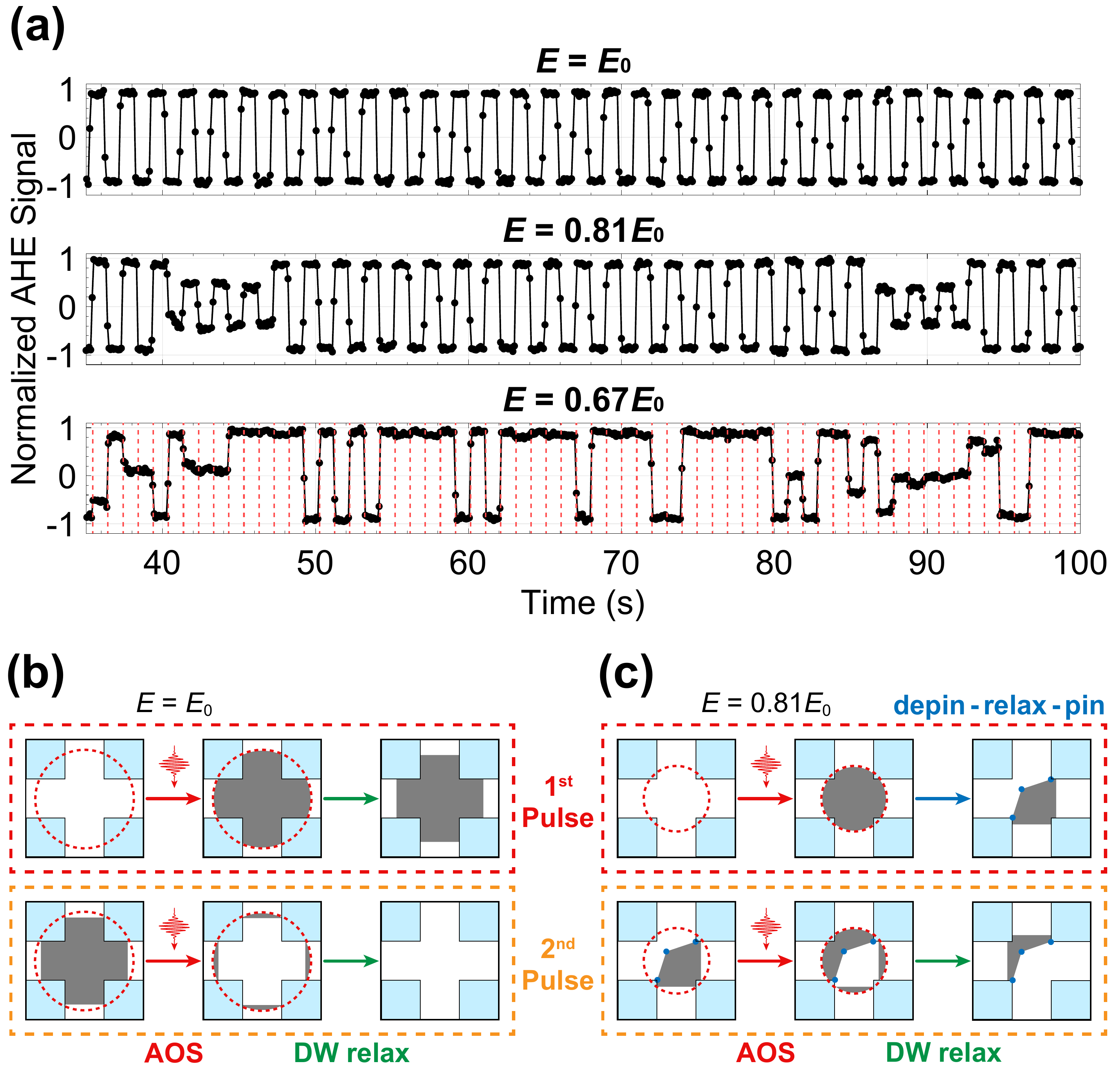}
    \caption{\textbf{Fig. 3 $\mid$ Near-threshold all-optical magnetization switching and domain wall dynamics.} \textbf{(a)} Time traces of normalized AHE signals for the 1~$\mathrm{\upmu}$m Hall cross device under front illumination at a 1~Hz repetition rate. Three pulse energies are shown: $E_{0}$, which slightly exceeds the switching threshold ($E_{\mathrm{th}}$) to achieve full magnetization reversal, $0.81 E_{0}$, and $0.67 E_{0}$. \textbf{(b,c)} Conceptual illustrations of domain wall dynamics under near-threshold illumination.}
    \label{fig:frontIllumination}
\end{figure}

We attribute this stochastic switching to the formation of meta-stable magnetic domain patterns (see S.I.~\textcolor{blue}{D}) within the Hall cross, resulting from a combination of toggle switching, domain wall (DW) relaxation, stochastic pinning and thermally assisted depinning events. Following magnetization reversal, a DW is formed and subsequently evolves toward a configuration that minimizes the total magnetic energy, including exchange energy, magnetocrystalline anisotropy energy, demagnetization energy, and possibly interfacial contributions such as the Dzyaloshinskii–Moriya interaction \cite{cao_dynamics_2020}. In the case of a 1~$\mathrm{\upmu}$m magnetic domain, the total energy is predominantly governed by the domain wall energy, which reflects the energy penalty from both exchange and anisotropy contributions due to the spatial rotation of magnetization across the wall. As a result, the DW tends to reduce its total energy by minimizing the domain perimeter, typically by straightening or retracting inward. However, this relaxation process can be impeded when the DW encounters structural boundaries or material defects, which can act as pinning sites \cite{bogart_dependence_2009} and prevent further energy minimization.

At pulse energies well above the switching threshold, the reversed domain extends across a large portion of the Hall cross (Fig.~\textcolor{blue}{3b}). During relaxation, the DW straightens and remains well-separated from the device corners, leading to a complete toggle of the AHE signal upon each laser pulse. Subsequent pulses erase the previous magnetic configuration, where any remaining newly generated magnetic domains due to pulse-to-pulse variations collapse upon relaxation, resulting in deterministic toggle switching.

At pulse energies near the switching threshold (e.g., at $0.81 E_{0}$), the DW can become pinned at sharp corners of the Hall cross \cite{bogart_dependence_2009} or trapped by local magnetic inhomogeneities (highlighted by blue circles in Fig.~\textcolor{blue}{3c}), most likely as a result of stray-field interactions. Residual heat from the laser pulse may occasionally lead to depinning, allowing the system to switch between complete and partial reversal states. In this regime, each laser pulse toggles the magnetization between two complementary configurations, but DW relaxation introduces stochasticity into the final state.

Such stochastic switching effects are expected to diminish as device dimensions approach or fall below the characteristic DW width. This observation underscores the promising scalability of integrated photonic control of AOS in nanoscale spintronic devices.

\section*{Scalability and integration prospects}\label{Sec: Discussion}
Having elucidated the origins of stochastic and partial switching under \textit{front illumination}, we now return to the more complex scenario of switching driven by guided photonic modes. Generally, partial and stochastic switching are considered detrimental for most practical applications, where deterministic and repeatable behavior is critical. However, it is important to note that when magnetic elements are scaled down to (sub-)100~nm dimensions, such stochasticity is largely suppressed due to the reduction in DW formation and pinning sites. Thus, device miniaturization inherently mitigates the challenges observed in larger devices.

That said, stochastic switching could also be exploited as a resource for certain probabilistic computing \cite{finocchio_promise_2021, incorvia_spintronics_2024} or neuromorphic applications \cite{grollier_neuromorphic_2020, marrows_neuromorphic_2024}, where controlled randomness may be advantageous. Future device designs could therefore be tailored to either suppress or harness this behavior, depending on the target functionality.

As mentioned previously, optimizing coupling efficiency was not the focus of the present work. For a preliminary estimate of coupling efficiencies, we refer the reader to the Methods section. Nonetheless, practical implementations could benefit significantly from improved coupling strategies, such as fiber-based approaches \cite{wang_scaling_2024} or the integration of on-chip pulsed laser sources \cite{gasse_recent_2019}, which would reduce the required input energy and enhance system efficiency.

In terms of operational speed, our proof-of-concept experiments employed a low repetition rate of 1~Hz for clarity and to facilitate detailed observation. However, prior studies have demonstrated the feasibility of much faster switching, with reported timescales down to 10~ps between pulses \cite{van_hees_toward_2022} and even sub-picosecond zero crossings \cite{li_picosecond_2025}. These findings suggest that ultrafast data processing based on integrated AOS is well within technological reach.

Concerning scalability (S.I.~\textcolor{blue}{E}) and integration potential, the dimensions of the magnetic element and those of the photonic routing layer need not scale simultaneously. In integrated photonic architectures, the waveguide primarily serves as a low-loss distribution network for optical signals, whereas sub-wavelength energy localization can be achieved locally through dedicated nanophotonic structures. To circumvent diffraction-limited optical confinement, our previous works have introduced photonic architectures based on plasmonic nanoantennas and photonic cavities that support sub-wavelength bit addressing below 100~nm {\cite{pezeshki_integrated_2023}}, while also enabling wavelength-division multiplexing for parallel data operations {\cite{pezeshki_integrated_2024}}. These results suggest that the practical scaling limit of photonic-assisted all-optical switching is determined less by the dimensions of the waveguide itself than by the efficiency of the optical-to-magnetic transduction mechanism and the stability of the nanoscale magnetic element. Furthermore, at deeply scaled dimensions, simple geometric downscaling of the Hall-cross architecture is unlikely to be optimal, as optical near-field effects can lead to increased parasitic absorption in surrounding device features, redirecting the optical energy away from the Hall-sensitive central region. Consequently, practical scaling to achieve refined on-chip deterministic AOS is expected to rely on co-designed optical interfaces and readout architectures that maximize energy delivery to the active magnetic region while minimizing unwanted excitation elsewhere in the device. Such architectures naturally extend the integrated photonic framework demonstrated here and provide a pathway toward technologically relevant device dimensions. Moreover, these architectures are directly compatible with the platform demonstrated here, as well as with straightforward extensions to MTJ readout {\cite{wang_picosecond_2022}}, racetrack memory {\cite{lalieu_deterministic_2017, pezeshki_integrated_2023}}, and magneto-photonic magnetization readout {\cite{demirer_integrated_2022}}, thereby providing clear pathways toward scalable, ultrafast spintronic-photonic hybrid systems.

\section*{Conclusions}\label{Sec: Conclusions}
In summary, we have demonstrated the realization of single-pulse all-optical switching (AOS) within a \ce{Si3N4}-based integrated photonic platform, bridging a critical gap between ultrafast spintronics and on-chip integrated photonics. By leveraging a generic Hall cross structure stacked on top of a \ce{Si3N4} waveguide, we achieved robust magnetization toggling in \ce{Co}/\ce{Gd} stacks under femtosecond laser excitation. While non-uniform light absorption in larger devices introduces challenges such as partial or stochastic switching, we confirmed that device miniaturization -- specifically, scaling below the characteristic domain wall (DW) width -- largely suppresses these effects.

These results not only demonstrate the feasibility of on-chip AOS but also lay the foundation for future advances in integrated spintronic-photonic hybridization. This includes the use of novel materials, alternative switching mechanisms, and advanced device and system architectures, all contributing to the development of energy-efficient and ultrafast data-processing platforms.

\backmatter
\bmhead{Acknowledgments}
The authors thank the technical staff of the NanoLabTUe for their continued support in maintaining the research facilities and instrumentation used throughout this work. We also thank J. Hintermayr for occasional technical assistance with the lock-in amplifier used during the experiments, T.J. Kools for initial technical support with the magnetron sputtering deposition system, and L. Gnoatto and A.A.D. Petrillo for valuable practical discussions on nanoscale patterning.

\bmhead{Funding statement}
This project received funding from the European Union’s Horizon 2020 research and innovation programme under the Marie Skłodowska-Curie grant agreement No. 860060. This work is also part of the Gravitation programme \textit{Research Centre for Integrated Nanophotonics}, financed by the Netherlands Organisation for Scientific Research (NWO), the PhotonDelta National Growth Fund programme, and the project \textit{NL-ECO: Netherlands Initiative for Energy-Efficient Computing} (project number NWA.1389.20.140) of the NWA research programme \textit{Research Along Routes by Consortia}, which is financed by the Dutch Research Council (NWO).

\bmhead{Author contributions}
P. Li conceived and designed the project. P. Li, G.W.A. Simons, and T. Zhang developed the experimental protocol, performed simulations, and collected data. G.W.A. Simons, P.P.J. Schrinner, and S. Kamyar were responsible for the fabrication of the devices. R. Dekker provided technical supervision of the photonic circuit, D.C. Leitao supervised the post-processing fabrication, and R. Lavrijsen and Y. Jiao provided feedback on the experimental work. P. Li and B. Koopmans supervised the project. P. Li wrote the original manuscript draft. All authors contributed to writing and editing the manuscript.

\bmhead{Competing interests}
The authors declare no competing interests.


\noindent\textbf{\large \\ Methods}\\ \vspace{-0.5cm}
\bmhead{Chip design and fabrication}
The integrated photonic platform was based on a silicon nitride (\ce{Si3N4}) waveguide system. Waveguides were fabricated using low-pressure chemical vapor deposition (LPCVD), ensuring high stoichiometry and low propagation loss \cite{roeloffzen_low-loss_2018}. A 8~$\mathrm{\upmu}$m thermally grown \ce{SiO2} layer served as the bottom cladding for optical isolation from the \ce{Si} substrate, while a 3.7~$\mathrm{\upmu}$m CVD \ce{SiO2} top cladding was deposited to minimize environmental contamination.

Inverse-tapered edge couplers were integrated at the chip periphery to enable efficient free-space coupling \cite{almeida_nanotaper_2003, roeloffzen_low-loss_2018, leinse_triplex_2013}. Optical characterization on test waveguides done by LioniX yielded an average fiber-to-fiber insertion loss of 6.00~dB and propagation loss of 0.68~dB/cm.

To allow optical interaction with the spintronic Hall cross, sensing windows were introduced by selectively etching the top \ce{SiO2} cladding, leaving a residual 50~nm layer in predefined regions. Lithographically defined \ce{Si3N4} stages, fabricated in the same step as the waveguides, provided a planar surface for Hall cross integration. The device topography was verified using optical microscopy, atomic force microscopy (AFM), and scanning electron microscopy (SEM).

\bmhead{Hall cross device fabrication}
The spintronic Hall cross was patterned by e-beam lithography and a lift-off process, and deposited directly onto the \ce{Si3N4} waveguide stage within the sensing window. The multilayer magnetic stack comprised \ce{Ta}(2)/\ce{Pt}(2)/\ce{Co}(1)/\ce{Gd}(3)/\ce{TaN}(5)/\ce{Pt}(2) (thickness within the parentheses are in nanometers), selected for its demonstrated compatibility with AOS. Hall crosses with various different arm widths were fabricated to align with the \ce{Si3N4} stage, enabling smooth electrical contact and efficient interactions. The central Hall cross section -- where magnetization switching occurs and is probed -- was aligned to the \ce{Si3N4} waveguide, and characterized using AFM and optical imaging.

In this work, we present results for $500 \times 500$~nm$^{2}$ and $1100 \times 700$~nm$^{2}$ Hall crosses. For the larger devices, portions of the arms overlap with the stage edges, reducing the effective electrical contact along the arms outside the central cross-section. However, because the stage center is formed by the intersection with the waveguide, the full diagonal of the Hall cross (length $\sqrt{700^{2} + 1100^{2}}$ nm) lies on top of the stage. We argue that this diagonal length governs the attenuation of the optical mode during propagation and is therefore critical to the occurrence of stochastic behavior.

\bmhead{Theoretical modeling and simulations}
Numerical simulations were performed using Ansys Lumerical Finite-Difference Time-Domain (FDTD) and Finite-Difference Eigenmode (FDE) solvers to evaluate the light absorption profile within the Hall cross. The occurrence of AOS in each cell was then assessed with a simplified M3TM model (see Refs. \cite{koopmans_explaining_2010, beens_comparing_2019} for details and parameter sets), using the relative differences in absorbed pulse energy from the FDTD output as input parameters. Dipolar spin interactions and other micromagnetic effects were not included, and thus the results in Fig.~\textcolor{blue}{2e} capture only the short-timescale dynamics of AOS. By contrast, the effects discussed in Fig.~\textcolor{blue}{3} occur on longer timescales, which are beyond the scope of our simulations.

\bmhead{Mode profile and propagation analysis}
The waveguide geometry was optimized to support only the fundamental transverse electric (TE) mode at 720~nm, matching the fs laser excitation wavelength. FDE simulations confirmed that the TE mode is weakly confined within the \ce{Si3N4} core due to the low refractive index contrast between \ce{Si3N4} ($n = 1.97$) and \ce{SiO2} ($n = 1.45$). This design enhances evanescent coupling into the spintronic stack, albeit at a slight cost to light confinement. FDTD simulations at the sensing window showed that over $93\%$ of the incoming optical power remains confined within the main waveguide, despite lateral discontinuities from the \ce{Si3N4} stages where light is expected to decouple into one of the other \ce{Si3N4} stage arms.

\bmhead{Absorption and switching thresholds}
FDTD simulations revealed spatially inhomogeneous absorption within a Ta/Pt/Co/Gd/Ta multilayer, with intensity decreasing along the direction of light propagation. Incorporating a 50~nm \ce{SiO2} spacer layer between the waveguide and Hall cross mitigated edge over-absorption. Furthermore, we note that increasing the spacer thickness up to 50~nm has no noteworthy effect on the absorption magnitude at the downstream edge of the Hall cross, thereby further improving the uniformity of the absorption profile. Rounding the corners of the Hall cross intersection produced a similar effect by reducing local plasmonic field enhancements near sharp metallic features. However, to avoid potential complications in measuring the AHE, rounded corners were not implemented. In this configuration, the total absorption within the Hall cross reached $\sim$41$\%$, obtained by summing the absorption across all FDTD cells covering the device.

\bmhead{Experimental setup}
A schematic of the experimental setup is shown in Fig.~\textcolor{blue}{1e}. Free-space, femtosecond laser pulses (720~nm wavelength) were focused onto the edge coupler using a high-numerical-aperture objective. In this respect, we note that the input coupling efficiency from free space into the waveguide was relatively low due to practical limitations and could be improved; however, in this conceptual study, optimizing coupling was not of primary importance. Stable, complete magnetization switching was first observed for optical pulses with an energy of $E_0=621$ nJ before entering the photonic circuit. A half-wave plate adjusted the polarization to match the TE mode of the \ce{Si3N4} waveguide. Beam alignment was aided by power detectors capturing reflected light, while optical neutral density filters regulated the pulse energy. A secondary beam could be directed perpendicularly onto the Hall cross for front-side illumination. 

During experiments, a current was supplied to one arm of the Hall cross and the anomalous Hall voltage was recorded along the orthogonal arm using a lock-in amplifier to allow precise electrical readout of the out-of-plane magnetization. The Hall cross was connected to this external equipment via wire bonding.

Because of the relatively large beam spot ($\sim$10~$\mathrm{\upmu}$m$^{2}$), along with suboptimal alignment and mode matching, the coupling efficiency was low, and changes in the reflected signal at the edge coupler were negligible.

\bmhead{Data availability}
The data supporting the findings of this study are available from the corresponding author upon request. The experimental procedures, data processing methods, and supporting datasets necessary to interpret and reproduce the reported results are provided in the Article and Supplementary Information.


\newpage

\begin{strip}
\begin{center}
\begin{minipage}{0.75\textwidth}
    \vspace{-1.3cm}
    \tableofcontents
    \vspace{0.5cm}
\end{minipage}
\end{center}
\end{strip}

\begin{appendices}
\section{Linear Drift Correction}\label{section:driftCorr}
To ensure transparency, Supplementary Fig.~\textcolor{blue}{1} presents the raw anomalous Hall effect (AHE) signal (blue curve), normalized to the full magnetization swing as determined from external magnetic-field sweeps. At $\sim$20 s, a femtosecond laser pulse train with a repetition rate of 1~Hz was applied in the absence of an external magnetic field, inducing toggle magnetization switching at the same rate via all-optical switching (AOS). During laser excitation, the AHE signal exhibits a gradual downward drift, which we attribute to localized absorption hot spots at the sharp edges of the Hall cross, leading to mild thermal degradation of the device. Importantly, this drift is non-magnetic in origin.

\begin{figure}[b]
    \centering
    \includegraphics[width=1\linewidth]{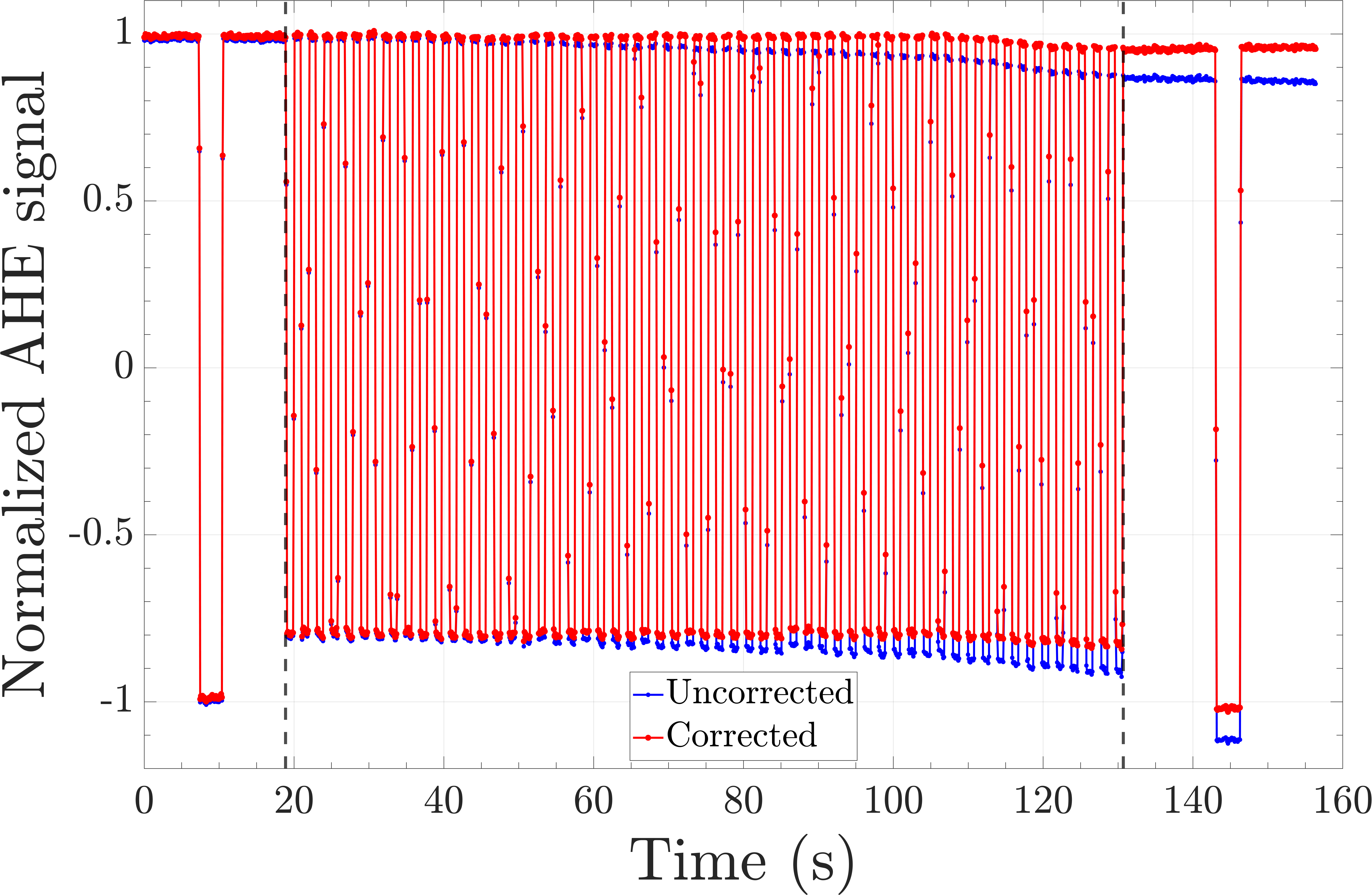}
    \caption{\small \textbf{Supplementary Fig. 1  $\mid$} Raw (blue) and drift-corrected (red) normalized anomalous Hall effect signal.}
    \label{fig:driftCorr}
\end{figure}

This interpretation is supported by external magnetic-field toggling performed both at the beginning ($\sim$15 s) and at the end ($\sim$140 s) of the measurement sequence without laser excitation. In both cases, the full magnetization swing is preserved, demonstrating that the intrinsic sensitivity of the Hall cross remains unchanged.

For clarity, a linear background correction was applied to the laser-excited data to remove the non-magnetic drift without altering the intrinsic toggle amplitude. The corrected signal (red curve) is shown alongside the raw data in Supplementary Fig.~\textcolor{blue}{1}, ensuring that the representation of laser-induced toggle switching is not obscured by unrelated thermal effects.

\section{Joined optical and magnetic model}\label{section:absorptionModel}
To better understand the conditions required to induce all-optical switching (AOS) in a magnetic Hall cross integrated on a \ce{Si3N4} waveguide, we developed a simplified model linking optical excitation to magnetization dynamics. The results are shown in Fig.~\textcolor{blue}{2e}. In ferrimagnetic materials, AOS critically depends on the absorbed optical power. To quantify this dependence, we employed the microscopic three-temperature model (M3TM) [\textcolor{blue}{19, 46}], which, despite its simplifications, captures the essential energy exchange among electrons, phonons, and spins following ultrafast excitation.

\begin{figure*}[t]
    \centering
    \includegraphics[width=0.8\linewidth]{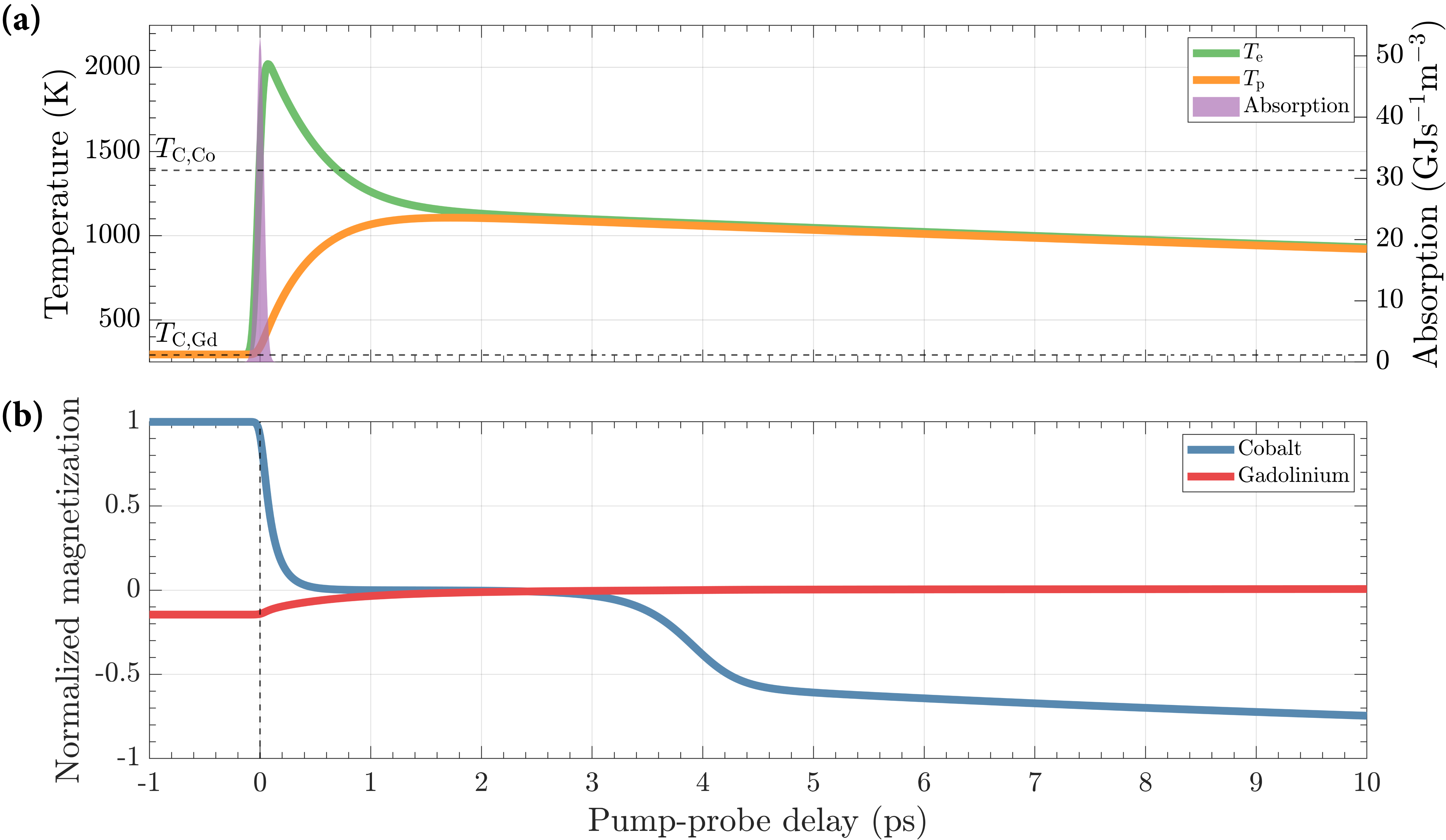}
    \caption{\small \textbf{Supplementary Fig. 2 $\mid$} Results of microscopic three-temperature model (M3TM) simulations for a \ce{Co}/\ce{Gd} system. \textbf{(a)} Temporal evolution of the electron ($T_{\mathrm{e}}$, blue) and phonon ($T_{\mathrm{p}}$, green) temperatures following optical excitation. The Gaussian pulse energy density (red) is derived from finite-difference time-domain (FDTD) simulations. Dashed lines indicate the Curie temperatures of cobalt ($T_{\mathrm{C,Co}} = 1388$~K) and gadolinium ($T_{\mathrm{C,Gd}} = 292$~K). \textbf{(b)} Corresponding relative magnetization dynamics of \ce{Co} (blue) and \ce{Gd} (red) on the same timescale, showing magnetization reversal in both layers with respect to their initial orientation.}
    \label{fig:2TM}
\end{figure*}

Optical absorption in the Hall cross was determined using finite-difference time-domain (FDTD) simulations. The local absorption density was calculated from the divergence of the Poynting vector as
\begin{equation}
    P_{\text{abs}}=\frac{1}{2} \omega \left\lvert E \right\rvert^{2} \text{Im}\left( \varepsilon \right),
    \label{eq:Pabs}
\end{equation}
where $\omega$ is the optical frequency, $E$ the electric-field amplitude, and $\text{Im}\left( \varepsilon \right)$ the imaginary component of the permittivity. The obtained absorption datasets were used as input for the M3TM, assuming a Gaussian temporal profile for the laser pulse.

For this study, the model was reduced to a two-temperature model (2TM) description, sufficient to capture the coupled electron–phonon temperature dynamics (Supplementary Fig.~\textcolor{blue}{2a}) that govern the final magnetic state (Supplementary Fig.~\textcolor{blue}{2b}), while reducing computational cost and complexity. The full M3TM was employed to determine the minimum peak electron temperature ($T_{\mathrm{e,max}}$) required to switch the magnetization of the \ce{Co}/\ce{Gd} system. This approach eliminates the need to recompute the full magnetization dynamics for each FDTD simulation cell, since the final magnetic state is uniquely determined by $T_{\mathrm{e,max}}$.

The energy exchange between the electron ($T_{\mathrm{e}}$) and phonon ($T_{\mathrm{p}}$) systems is described by
\begin{equation}
\begin{aligned}
    \gamma T_{\mathrm{e}} \frac{dT_{\mathrm{e}}}{dt} &= g_{\mathrm{ep}}(T_{\mathrm{p}} - T_{\mathrm{e}}) + P(t) \\
    C_{\mathrm{p}} \frac{dT_{\mathrm{p}}}{dt} &= g_{\mathrm{ep}}(T_{\mathrm{e}} - T_{\mathrm{p}}) + C_{\mathrm{p}} \frac{(T_{\mathrm{amb}} - T_{\mathrm{p}})}{\tau_{\mathrm{d}}},
\end{aligned}
\label{eq:2TM}
\end{equation}
where $\gamma$ is the electronic heat-capacity constant, $C_{\mathrm{p}}$ the phonon heat capacity (Debye model), $g_{\mathrm{ep}}$ the electron–phonon coupling constant, $\tau_{\mathrm{d}}$ the characteristic heat diffusion time, and $T_{\mathrm{amb}}$ the ambient temperature. The laser power $P\left(t\right)$ follows a Gaussian temporal profile centered at $t_0 = 4\sigma$ with standard deviation $\sigma = 50$~fs, and its amplitude $P_0$ is scaled to the absorbed power density $P_{\text{abs}}$ from Eq.~(\textcolor{blue}{B1}).

As shown in Supplementary Fig.~\textcolor{blue}{2}, the electron temperature ($T_{\mathrm{e}}$, blue) rises sharply following excitation (red), followed by a slower phonon response ($T_{\mathrm{p}}$, green) due to energy transfer. After thermal equilibrium is reached, the system cools through phonon-mediated heat diffusion to the substrate.

AOS occurs when the peak electron temperature ($T_{\mathrm{e,max}}$) exceeds the Curie temperature of cobalt ($T_{\mathrm{C,Co}} = 1388$~K) while the lattice temperature remains below it ($T_{\mathrm{e}} > T_{\mathrm{C,Co}} > T_{\mathrm{p}}$). It should be noted that in magnetic thin films, the Curie temperature is typically much lower than the corresponding bulk value. Although the exact value is challenging to determine experimentally, using the bulk Curie temperature in these calculations remains a reasonable approximation, as variations in $T_\mathrm{C}$ have only a minor influence on the relative magnetization trends observed. For the studied multilayer, \ce{Ta}(2)/\ce{Pt}(2)/\ce{Co}(1)/\ce{Gd}(3)/\ce{TaN}(5)/\ce{Pt}(2), where numbers in parentheses denote layer thicknesses in nanometers, simulations show that switching occurs when $T_{\mathrm{e,max}} > 2020$~K~$> T_{\mathrm{p,max}}$. The corresponding temperature and magnetization dynamics are shown in Supplementary Fig.~\textcolor{blue}{2}. At higher excitation levels where $T_{\mathrm{p}} > T_{\mathrm{C,Co}}$, complete demagnetization occurs, and remagnetization becomes stochastic, resulting in a multidomain state. These temperatures, based on bulk Curie values, are expected to exceed those experimentally observed in thin films.

\section{On-chip AOS under varied input energies}\label{section:extraData}
We examined the on-chip all-optical switching (AOS) behavior following the \textit{edge illumination} protocol for a range of pulse energies beyond the reference case of 621 nJ ($E_0$, Fig.~\textcolor{blue}{2a}) and 527 nJ (Fig.~\textcolor{blue}{2d}). For the same Hall cross device, Supplementary Fig.~\textcolor{blue}{3} presents the time traces of the anomalous Hall voltage for optical excitation (between the red dashed lines) at pulse energies of \textbf{(a)} 586 nJ, \textbf{(b)} 558 nJ, and \textbf{(c)} 502 nJ. All traces are normalized to the initial magnetization swing induced by an external magnetic field which are also present in the figures.

\begin{figure}[t]
    \centering
    \includegraphics[width=1\linewidth]{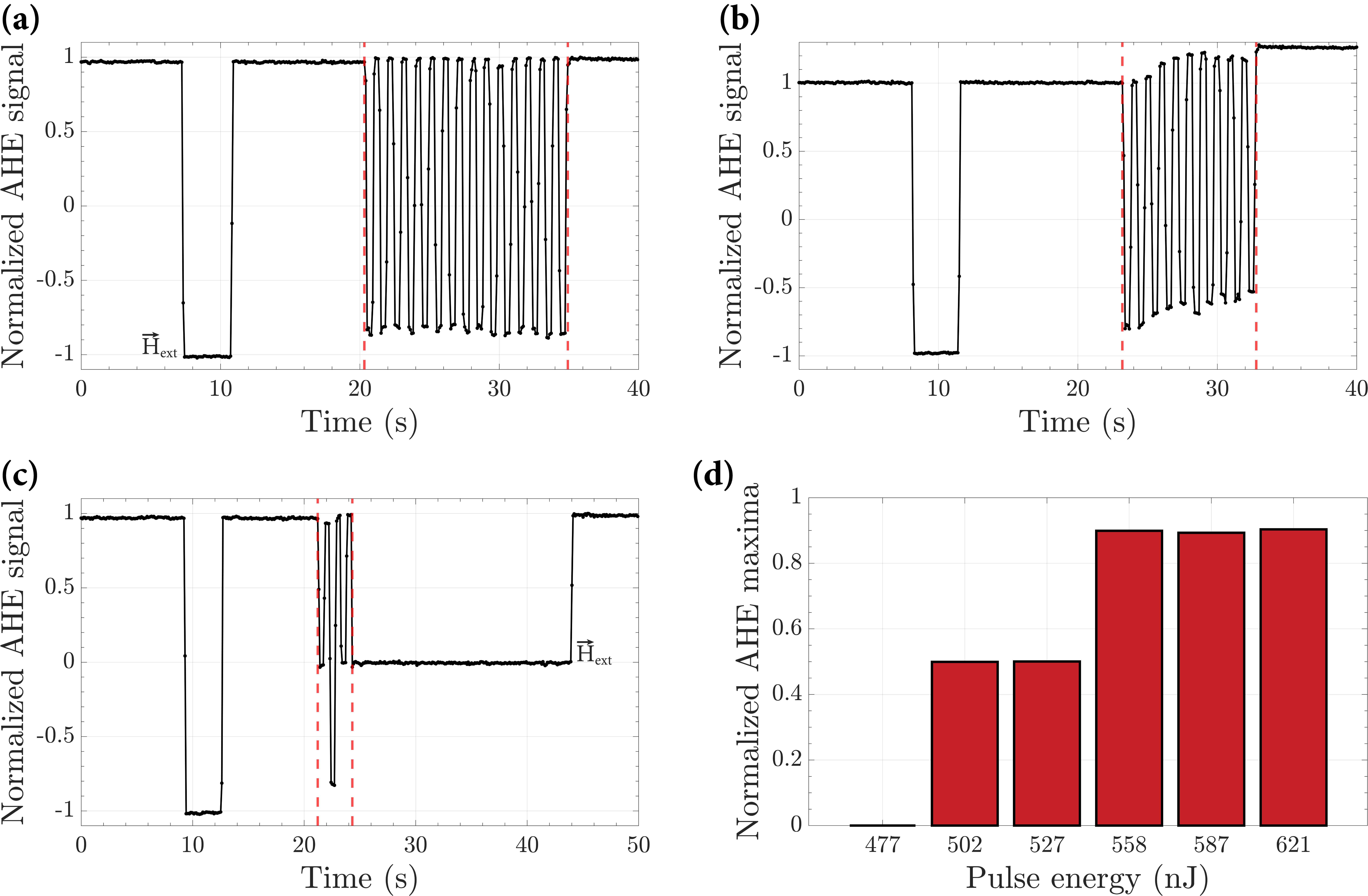}
    \caption{\small \textbf{Supplementary Fig. 3 $\mid$} On-chip all-optical magnetization switching at different pulse energies. \textbf{(a–c)} Time traces of the anomalous Hall voltage recorded at pulse energies of 586~nJ, 558~nJ, and 502~nJ, respectively. The first two reversals are induced by an external magnetic field to define the saturation magnetization, to which all subsequent measurements are normalized. Between the red dashed lines, the magnetic field is removed and switching occurs purely optically under femtosecond light pulses at a repetition rate of 2~Hz. \textbf{(d)} Average fraction of the Hall cross area switched as a function of pulse energy, accounting for stochastic variations at lower energies.}
    \label{fig:ExtraAOS}
\end{figure}

As discussed in the main text, reducing the pulse energy below the stable-switching threshold prevents complete reversal of the Hall cross magnetization. Instead, the system stabilizes in a pinned intermediate state, where approximately half of the area probed by the anomalous Hall effect is switched. This behavior is particularly evident in Supplementary Fig.~\textcolor{blue}{3c}, where the intermediate state remains stable between 25~s and 44~s until an external magnetic field is reapplied.

The average switched area was estimated by comparing the total anomalous Hall voltage change induced by the optical pulses to the voltage contrast obtained from full magnetization reversal under an external field. The corresponding results, shown in Supplementary Fig.~\textcolor{blue}{3d}, indicate that the threshold pulse energy for stable complete switching of the Hall cross lies between 527~nJ and 558~nJ. At a pulse energy of 477~nJ, no AOS is observed.

\section{Visualization of magnetic domains}\label{section:MFM}
To further support the hypothesis that magnetic domain wall (DW) relaxation plays a role in predicting the final magnetic configuration of the Hall cross, as described in Fig.~\textcolor{blue}{3b}, we experimentally visualized a partially switched magnetic state using magnetic force microscopy (MFM). The measurements, shown in Supplementary Fig.~\textcolor{blue}{4}, were performed after \textit{front-illumination} excitation with a 5.5~$\mathrm{\upmu}$m laser spot on a magnetic Hall cross (not integrated with the photonic platform) featuring 500~nm-wide arms.

\begin{figure}[t]
    \centering
    \includegraphics[width=1\linewidth]{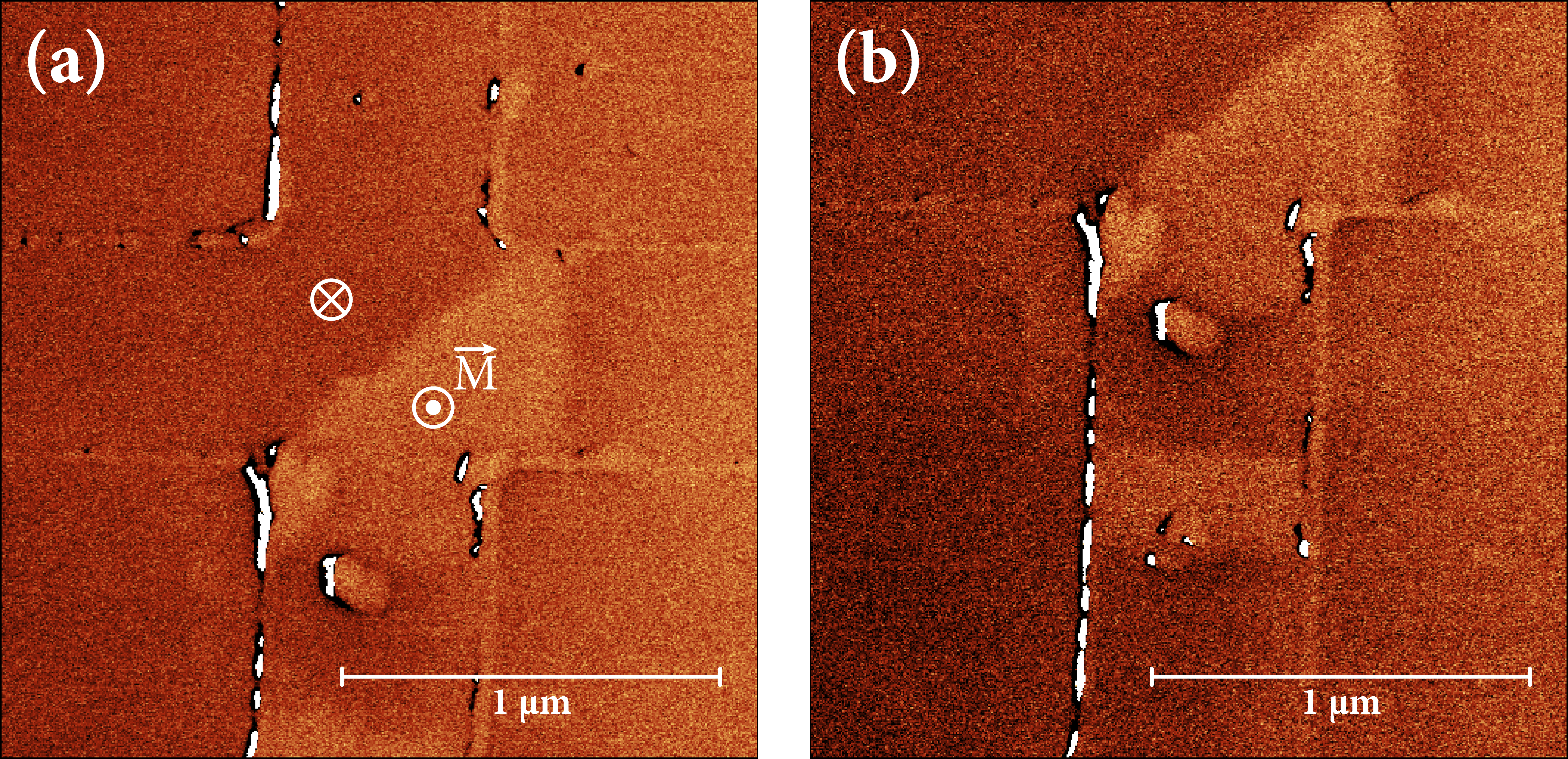}
    \caption{\small \textbf{Supplementary Fig. 4 $\mid$} Magnetic force microscopy of the Hall cross. \textbf{(a)} Central region and \textbf{(b)} one arm of the Hall cross after \textit{front-illumination} excitation. Variations in orange hue correspond to opposite out-of-plane magnetization orientations as labeled in (a). The observed magnetic patterns indicate magnetic domain wall pinning at sharp corners and relaxation.}
    \label{fig:MFM}
\end{figure}

For these measurements, an atomic force microscopy (AFM) tip was used, featuring a resonant frequency of 75~kHz, a cantilever spring constant of 2.8~N/m, and a tip radius of curvature below 10~nm. On top of the standard aluminium coating, an additional magnetic heterostructure of \ce{Ta}(4)/\ce{Co}(5)/\ce{Ta}(4) (thicknesses in nanometers) was deposited on the cantilever by magnetron sputtering. The material of the Hall cross studied here differs slightly from those integrated into the photonic circuit and consists of \ce{Ta}(4)/\ce{Pt}(4)/\ce{Co}(0.6)/\ce{Gd}(1.2)/\ce{Co}(0.6)/\ce{Pt}(1.25)/ \ce{Co}(0.6)/\ce{Pt}(4), providing enhanced magnetic contrast owing to its higher overall magnetic moment.

MFM imaging was performed in lifting mode, in which the Hall cross surface topography is first acquired in tapping mode and subsequently retraced at a fixed lift height to sense the long-range magnetic interaction between the tip and the sample stray fields. Basic post-processing, such as adjusting the colormap height range, was applied to enhance the visible magnetic contrast in the resulting images.

As shown in Supplementary Fig.~\textcolor{blue}{4a}, the MFM image visualizes a partially switched magnetic state. The incident laser power was deliberately reduced to introduce stochasticity into the switching process, and the excitation was stopped once the anomalous Hall effect (AHE) signal indicated a partial reversal. A straight DW is clearly observed between oppositely magnetized regions, extending diagonally from the lower left to the upper right corner of the Hall cross. This configuration supports the proposed mechanism of magnetic relaxation and DW pinning at sharp geometric discontinuities. In the absence of apparent defects, the DW minimizes its total energy by straightening and becomes pinned only at the two corners of the Hall cross. In Supplementary Fig.~\textcolor{blue}{4b}, an additional magnetic domain appears in one of the Hall cross arms, consistent with the excitation spot slightly exceeding the central region. The final domain adopts a rectangular shape, further supporting the proposed mechanism that magnetic relaxation favors straight DWs, minimizing the DW circumference, constrained by the device geometry.

\section{Estimate of the switching energy}\label{section:CouplingEfficiency}
To estimate the switching threshold energy at scaled magnetic elements, we combine experimentally measured free-space all-optical switching (AOS) thresholds with finite-difference time-domain (FDTD) simulations of optical absorption.

\begin{figure}[t]
    \centering
    \includegraphics[width=1\linewidth]{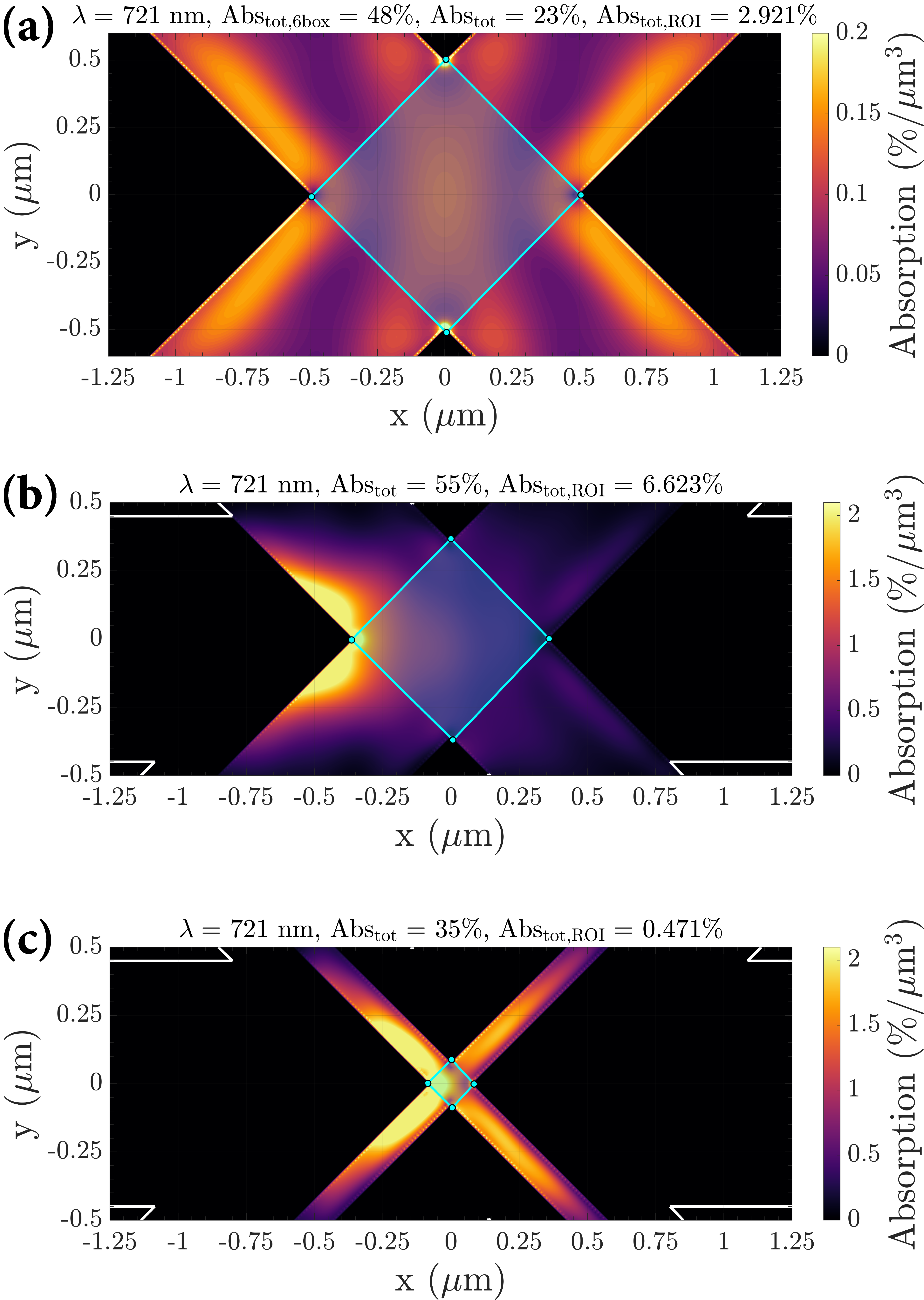}
    \caption{\small \textbf{Supplementary Fig. 5 $\mid$} Simulated optical absorption and scaling behavior of Hall-cross devices under free-space and waveguide excitation. Top-view absorption maps obtained from finite-difference time-domain (FDTD) simulations. The Hall-signal-sensitive region of interest (ROI) is highlighted in blue. {\textbf{(a)}} A 700~nm Hall-cross device under free-space excitation, modeled as a plane wave polarized along the {$y$}-direction. {\textbf{(b)}} A 500~nm Hall-cross device excited by the fundamental transverse-electric (TE) mode of a {$900 {\times} 100$} nm{$^2$} {\ce{Si3N4}} waveguide. {\textbf{(c)}} A scaled 100~nm Hall-cross device under the same waveguide excitation conditions. The absorption fractions within the ROI and the entire Hall cross are indicated in each panel, illustrating how the overlap between the optical mode and the magnetic element evolves with device scaling. For panel {\textbf{(a)}}, the absorption calculated via a transmission box surrounding the device is additionally reported.}
    \label{fig:OpticalAbsorption}
\end{figure}

Experimental free-space AOS measurements on a {\ce{Co}/\ce{Gd}} Hall-cross device with a 700~nm arm width yield a threshold switching fluence of approximately 5~mJ/cm{$^2$}. FDTD simulations of the optical excitation (Supplementary Fig.~{\textcolor{blue}{5a}}), modeled as a plane wave to reflect the experimental condition where the laser spot size is much larger than the device dimensions, indicate that approximately 16\% of the regional incident optical energy is absorbed within the Hall-sensitive region (highlighted in blue). This corresponds to an absorbed switching fluence of approximately 0.8~mJ/cm{$^2$}.

Assuming that switching is solely governed by the absorbed optical energy density within the Hall-sensitive region and scales linearly, the corresponding absorbed switching energy for a 500~nm Hall-cross device is estimated to be approximately 2.1~pJ. We then use FDTD simulations of the integrated photonic structure to determine the coupling efficiency from the fundamental transverse-electric (TE) waveguide mode into the Hall-sensitive region, visualized in Supplementary Fig.~{\textcolor{blue}{5b}}. For the present device geometry, approximately 6.6\% of the optical energy propagating in the waveguide is absorbed within the Hall-sensitive region (region-of-interest (ROI)). Consequently, the required in-waveguide pulse energy to reach the estimated switching threshold is approximately 32~pJ.

To assess the scaling behavior of the concept, we performed the same analysis for a 100~nm Hall-cross device while maintaining the same waveguide geometry. Owing to a more homogeneous absorption distribution within the active region, visible in Supplementary Fig.~{\textcolor{blue}{5c}}, the required pulse energy is reduced to approximately 18~pJ. However, the improvement is significantly smaller than the geometric area reduction because the optical mode remains substantially larger than the interaction cross-section, limiting the scaling of the optical coupling efficiency.

These estimates should be regarded as first-order values intended to quantify the energy requirements of the present proof-of-concept architecture. More advanced optical confinement schemes, such as photonic-crystal or plasmonic nanofocusing structures {[\textcolor{blue}{32}]}, are expected to substantially increase the fraction of optical energy delivered to the active region and thereby further reduce the switching input energy.

\end{appendices}


\begin{thebibliography}{10}

\bibitem{dieny_opportunities_2020}
B.~Dieny, I.~L. Prejbeanu, K.~Garello, P.~Gambardella, P.~Freitas, R.~Lehndorff, W.~Raberg, U.~Ebels, S.~O. Demokritov, J.~Akerman, A.~Deac, P.~Pirro, C.~Adelmann, A.~Anane, A.~V. Chumak, A.~Hirohata, S.~Mangin, S.~O. Valenzuela, M.~C. Onbaşlı, M.~d’Aquino, G.~Prenat, G.~Finocchio, L.~Lopez-Diaz, R.~Chantrell, O.~Chubykalo-Fesenko, and P.~Bortolotti, ``Opportunities and challenges for spintronics in the microelectronics industry,'' {\em Nat. Electron.}, vol.~3, no.~8, pp.~446--459, 2020.

\bibitem{nguyen_recent_2024}
V.~D. Nguyen, S.~Rao, K.~Wostyn, and S.~Couet, ``Recent progress in spin-orbit torque magnetic random-access memory,'' {\em npj Spintronics}, vol.~2, no.~1, p.~48, 2024.

\bibitem{mishra_voltage-controlled_2024}
P.~K. Mishra, M.~Sravani, A.~Bose, and S.~Bhuktare, ``Voltage-controlled magnetic anisotropy-based spintronic devices for magnetic memory applications: {Challenges} and perspectives,'' {\em J. Appl. Phys.}, vol.~135, no.~22, p.~220701, 2024.

\bibitem{parkin_memory_2015-1}
S.~Parkin and S.-H. Yang, ``Memory on the racetrack,'' {\em Nat. Nanotechnol.}, vol.~10, no.~3, pp.~195--198, 2015.
\newblock Publisher: Nature Publishing Group.

\bibitem{ralph_spin_2008}
D.~Ralph and M.~Stiles, ``Spin transfer torques,'' {\em J. Magn. Magn. Mater.}, vol.~320, no.~7, pp.~1190--1216, 2008.

\bibitem{choe_recent_2023}
J.~Choe, ``Recent {Technology} {Insights} on {STT}-{MRAM}: {Structure}, {Materials}, and {Process} {Integration},'' in {\em 2023 {IEEE} {International} {Memory} {Workshop} ({IMW})}, pp.~1--4, 2023.

\bibitem{beaurepaire_ultrafast_1996}
E.~Beaurepaire, J.-C. Merle, A.~Daunois, and J.-Y. Bigot, ``Ultrafast spin dynamics in ferromagnetic nickel,'' {\em Phys. Rev. Lett.}, vol.~76, no.~22, pp.~4250--4253, 1996.

\bibitem{walowski_perspective_2016}
J.~Walowski and M.~Münzenberg, ``Perspective: {Ultrafast} magnetism and {THz} spintronics,'' {\em J. Appl. Phys.}, vol.~120, p.~140901, Aug. 2016.

\bibitem{kimel_writing_2019}
A.~V. Kimel and M.~Li, ``Writing magnetic memory with ultrashort light pulses,'' {\em Nat. Rev. Mater.}, vol.~4, no.~3, pp.~189--200, 2019.

\bibitem{bull_spintronic_2021}
C.~Bull, S.~M. Hewett, R.~Ji, C.-H. Lin, T.~Thomson, D.~M. Graham, and P.~W. Nutter, ``Spintronic terahertz emitters: Status and prospects from a materials perspective,'' {\em APL Materials}, vol.~9, p.~090701, 09 2021.

\bibitem{wang_picosecond_2022}
L.~Wang, H.~Cheng, P.~Li, Y.~L.~W. van Hees, Y.~Liu, K.~Cao, R.~Lavrijsen, X.~Lin, B.~Koopmans, and W.~Zhao, ``Picosecond optospintronic tunnel junctions,'' {\em Proceedings of the National Academy of Sciences}, vol.~119, no.~24, p.~e2204732119, 2022.

\bibitem{Salomoni_field_2023}
D.~Salomoni, Y.~Peng, L.~Farcis, S.~Auffret, M.~Hehn, G.~Malinowski, S.~Mangin, B.~Dieny, L.~Buda-Prejbeanu, R.~Sousa, and I.~Prejbeanu, ``Field-free all-optical switching and electrical readout of $\mathrm{Tb}$/$\mathrm{Co}$-based magnetic tunnel junctions,'' {\em Phys. Rev. Appl.}, vol.~20, p.~034070, Sep 2023.

\bibitem{yang_ultrafast_2017}
Y.~Yang, R.~B. Wilson, J.~Gorchon, C.-H. Lambert, S.~Salahuddin, and J.~Bokor, ``Ultrafast magnetization reversal by picosecond electrical pulses,'' {\em Sci. Adv.}, vol.~3, no.~11, p.~e1603117, 2017.

\bibitem{polley_picosecond_2023}
D.~Polley, A.~Pattabi, A.~Rastogi, K.~Jhuria, E.~Diaz, H.~Singh, A.~Lemaitre, M.~Hehn, J.~Gorchon, and J.~Bokor, ``Picosecond spin-orbit torque–induced coherent magnetization switching in a ferromagnet,'' {\em Sci. Adv.}, vol.~9, p.~eadh5562, Sept. 2023.

\bibitem{stanciu_all-optical_2007}
C.~D. Stanciu, F.~Hansteen, A.~V. Kimel, A.~Kirilyuk, A.~Tsukamoto, A.~Itoh, and T.~Rasing, ``All-optical magnetic recording with circularly polarized light,'' {\em Phys. Rev. Lett.}, vol.~99, no.~4, p.~047601, 2007.

\bibitem{ostler_ultrafast_2012}
T.~A. Ostler, J.~Barker, R.~F.~L. Evans, R.~W. Chantrell, U.~Atxitia, O.~Chubykalo-Fesenko, S.~El~Moussaoui, L.~Le~Guyader, E.~Mengotti, L.~J. Heyderman, F.~Nolting, A.~Tsukamoto, A.~Itoh, D.~Afanasiev, B.~A. Ivanov, A.~M. Kalashnikova, K.~Vahaplar, J.~Mentink, A.~Kirilyuk, T.~Rasing, and A.~V. Kimel, ``Ultrafast heating as a sufficient stimulus for magnetization reversal in a ferrimagnet,'' {\em Nat. Commun.}, vol.~3, no.~1, p.~666, 2012.

\bibitem{zhang_paradigm_2009}
G.~P. Zhang, W.~Hübner, G.~Lefkidis, Y.~Bai, and T.~F. George, ``Paradigm of the time-resolved magneto-optical kerr effect for femtosecond magnetism,'' {\em Nat. Phys.}, vol.~5, no.~7, pp.~499--502, 2009.

\bibitem{demirer_integrated_2022}
F.~E. Demirer, Y.~Baron, S.~Reniers, D.~Pustakhod, R.~Lavrijsen, J.~v.~d. Tol, and B.~Koopmans, ``An integrated photonic device for on-chip magneto-optical memory reading,'' {\em Nanophotonics}, vol.~11, no.~14, pp.~3319--3329, 2022.

\bibitem{koopmans_explaining_2010}
B.~Koopmans, G.~Malinowski, F.~Dalla~Longa, D.~Steiauf, M.~Fähnle, T.~Roth, M.~Cinchetti, and M.~Aeschlimann, ``Explaining the paradoxical diversity of ultrafast laser-induced demagnetization,'' {\em Nat. Mater.}, vol.~9, no.~3, pp.~259--265, 2010.

\bibitem{radu_transient_2011}
I.~Radu, K.~Vahaplar, C.~Stamm, T.~Kachel, N.~Pontius, H.~A. Dürr, T.~A. Ostler, J.~Barker, R.~F.~L. Evans, R.~W. Chantrell, A.~Tsukamoto, A.~Itoh, A.~Kirilyuk, T.~Rasing, and A.~V. Kimel, ``Transient ferromagnetic-like state mediating ultrafast reversal of antiferromagnetically coupled spins,'' {\em Nature}, vol.~472, no.~7342, pp.~205--208, 2011.

\bibitem{alebrand_interplay_2012}
S.~Alebrand, A.~Hassdenteufel, D.~Steil, M.~Cinchetti, and M.~Aeschlimann, ``Interplay of heating and helicity in all-optical magnetization switching,'' {\em Phys. Rev. B}, vol.~85, no.~9, p.~092401, 2012.
\newblock Publisher: American Physical Society.

\bibitem{parlak_optically_2018}
U.~Parlak, R.~Adam, D.~E. B\"urgler, S.~Gang, and C.~M. Schneider, ``Optically induced magnetization reversal in ${[\mathrm{Co}/\mathrm{Pt}]}_{N}$ multilayers: Role of domain wall dynamics,'' {\em Phys. Rev. B}, vol.~98, p.~214443, Dec 2018.

\bibitem{van_hees_toward_2022}
Y.~L.~W. van Hees, B.~Koopmans, and R.~Lavrijsen, ``Toward high all-optical data writing rates in synthetic ferrimagnets,'' {\em Appl. Phys. Lett.}, vol.~120, no.~25, p.~252401, 2022.

\bibitem{wang_enhanced_2020}
L.~Wang, Y.~L.~W. van Hees, R.~Lavrijsen, W.~Zhao, and B.~Koopmans, ``Enhanced all-optical switching and domain wall velocity in annealed synthetic-ferrimagnetic multilayers,'' {\em Appl. Phys. Lett.}, vol.~117, p.~022408, 07 2020.

\bibitem{li_ultra-low_2021}
P.~Li, M.~J.~G. Peeters, Y.~L.~W. van Hees, R.~Lavrijsen, and B.~Koopmans, ``Ultra-low energy threshold engineering for all-optical switching of magnetization in dielectric-coated co/gd based synthetic-ferrimagnet,'' {\em Appl. Phys. Lett.}, vol.~119, no.~25, p.~252402, 2021.

\bibitem{wei_all-optical_2021}
J.~Wei, B.~Zhang, M.~Hehn, W.~Zhang, G.~Malinowski, Y.~Xu, W.~Zhao, and S.~Mangin, ``All-optical helicity-independent switching state diagram in gd-fe-co alloys,'' {\em Phys. Rev. Appl.}, vol.~15, no.~5, p.~054065, 2021.

\bibitem{peng_in-plane_2023}
Y.~Peng, D.~Salomoni, G.~Malinowski, W.~Zhang, J.~Hohlfeld, L.~D. Buda-Prejbeanu, J.~Gorchon, M.~Vergès, J.~X. Lin, D.~Lacour, R.~C. Sousa, I.~L. Prejbeanu, S.~Mangin, and M.~Hehn, ``In-plane reorientation induced single laser pulse magnetization reversal,'' {\em Nat. Commun.}, vol.~14, no.~1, p.~5000, 2023.

\bibitem{verges_extending_2024}
M.~Verges, W.~Zhang, Q.~Remy, Y.~Le-Guen, J.~Gorchon, G.~Malinowski, S.~Mangin, M.~Hehn, and J.~Hohlfeld, ``Extending the scope and understanding of all-optical magnetization switching in gd-based alloys by controlling the underlying temperature transients,'' {\em Phys. Rev. Appl.}, vol.~21, no.~4, p.~044003, 2024.

\bibitem{li_picosecond_2025}
P.~Li, T.~J. Kools, H.~Pezeshki, J.~M. B.~E. Joosten, J.~Li, J.~Igarashi, J.~Hohlfeld, R.~Lavrijsen, S.~Mangin, G.~Malinowski, and B.~Koopmans, ``Picosecond all-optical switching of co/gd--based synthetic ferrimagnets,'' {\em Phys. Rev. B}, vol.~111, p.~064421, Feb 2025.

\bibitem{lalieu_integrating_2019}
M.~L.~M. Lalieu, R.~Lavrijsen, and B.~Koopmans, ``Integrating all-optical switching with spintronics,'' {\em Nat. Commun.}, vol.~10, no.~1, p.~110, 2019.

\bibitem{li_ultrafast_2023}
P.~Li, T.~J. Kools, B.~Koopmans, and R.~Lavrijsen, ``Ultrafast racetrack based on compensated co/gd-based synthetic ferrimagnet with all-optical switching,'' {\em Adv. Electron. Mater.}, vol.~9, no.~1, p.~2200613, 2023.

\bibitem{pezeshki_integrated_2023}
H.~Pezeshki, P.~Li, R.~Lavrijsen, M.~Heck, E.~Bente, J.~van~der Tol, and B.~Koopmans, ``Integrated hybrid plasmonic-photonic device for all-optical switching and reading of spintronic memory,'' {\em Phys. Rev. Appl.}, vol.~19, no.~5, p.~054036, 2023.

\bibitem{siew_review_2021}
S.~Y. Siew, B.~Li, F.~Gao, H.~Y. Zheng, W.~Zhang, P.~Guo, S.~W. Xie, A.~Song, B.~Dong, L.~W. Luo, C.~Li, X.~Luo, and G.-Q. Lo, ``Review of silicon photonics technology and platform development,'' {\em Journal of Lightwave Technology}, vol.~39, no.~13, pp.~4374--4389, 2021.

\bibitem{wang_scaling_2024}
Y.~Wang, Y.~Jiao, and K.~Williams, ``Scaling photonic integrated circuits with inp technology: A perspective,'' {\em APL Photonics}, vol.~9, p.~050902, 05 2024.

\bibitem{miller_integrated_1969}
S.~E. Miller, ``Integrated optics: An introduction,'' {\em Bell System Technical Journal}, vol.~48, no.~7, pp.~2059--2069, 1969.

\bibitem{becker_out_2019}
H.~Becker, C.~J. Krückel, D.~Van~Thourhout, and M.~J.~R. Heck, ``Out-of-plane focusing grating couplers for silicon photonics integration with optical mram technology,'' {\em IEEE J. Sel. Top. Quantum Electron.}, vol.~26, no.~2, pp.~1--8, 2020.

\bibitem{alexoudi_optical_2020}
T.~Alexoudi, G.~T. Kanellos, and N.~Pleros, ``Optical {RAM} and integrated optical memories: a survey,'' {\em Light Sci. Appl.}, vol.~9, no.~1, p.~91, 2020.

\bibitem{youngblood_integrated_2023}
N.~Youngblood, C.~A. Ríos~Ocampo, W.~H.~P. Pernice, and H.~Bhaskaran, ``Integrated optical memristors,'' {\em Nat. Photon.}, vol.~17, no.~7, pp.~561--572, 2023.

\bibitem{pezeshki_integrated_2024}
H.~Pezeshki, P.~Li, R.~Lavrijsen, M.~Heck, and B.~Koopmans, ``Integrated magneto-photonic non-volatile multi-bit memory,'' {\em J. Appl. Phys.}, vol.~136, no.~8, p.~083908, 2024.

\bibitem{roeloffzen_low-loss_2018}
C.~G.~H. Roeloffzen, M.~Hoekman, E.~J. Klein, L.~S. Wevers, R.~B. Timens, D.~Marchenko, D.~Geskus, R.~Dekker, A.~Alippi, R.~Grootjans, A.~van Rees, R.~M. Oldenbeuving, J.~P. Epping, R.~G. Heideman, K.~Wörhoff, A.~Leinse, D.~Geuzebroek, E.~Schreuder, P.~W.~L. van Dijk, I.~Visscher, C.~Taddei, Y.~Fan, C.~Taballione, Y.~Liu, D.~Marpaung, L.~Zhuang, M.~Benelajla, and K.-J. Boller, ``Low-loss si3n4 {TriPleX} optical waveguides: Technology and applications overview,'' {\em IEEE J. Sel. Top. Quantum Electron.}, vol.~24, no.~4, pp.~1--21, 2018.

\bibitem{kaur_hybrid_2021}
P.~Kaur, A.~Boes, G.~Ren, T.~G. Nguyen, G.~Roelkens, and A.~Mitchell, ``Hybrid and heterogeneous photonic integration,'' {\em APL Photonics}, vol.~6, no.~6, p.~061102, 2021.

\bibitem{lalieu_deterministic_2017}
M.~L.~M. Lalieu, M.~J.~G. Peeters, S.~R.~R. Haenen, R.~Lavrijsen, and B.~Koopmans, ``Deterministic all-optical switching of synthetic ferrimagnets using single femtosecond laser pulses,'' {\em Phys. Rev. B}, vol.~96, no.~22, p.~220411, 2017.

\bibitem{kools_aging_2023}
T.~J. Kools, Y.~L.~W. van Hees, K.~Poissonnier, P.~Li, B.~Barcones~Campo, M.~A. Verheijen, B.~Koopmans, and R.~Lavrijsen, ``Aging and passivation of magnetic properties in co/gd bilayers,'' {\em Appl. Phys. Lett.}, vol.~123, no.~4, p.~042406, 2023.

\bibitem{almeida_nanotaper_2003}
V.~R. Almeida, R.~R. Panepucci, and M.~Lipson, ``Nanotaper for compact mode conversion,'' {\em Opt. Lett.}, vol.~28, pp.~1302--1304, Aug 2003.

\bibitem{leinse_triplex_2013}
A.~Leinse, R.~G. Heideman, M.~Hoekman, F.~Schreuder, F.~Falke, C.~G.~H. Roeloffzen, L.~Zhuang, M.~Burla, D.~Marpaung, D.~H. Geuzebroek, R.~Dekker, E.~J. Klein, P.~W.~L. van Dijk, and R.~M. Oldenbeuving, ``{TriPleX waveguide platform: low-loss technology over a wide wavelength range},'' in {\em Integrated Photonics: Materials, Devices, and Applications II} (J.-M. F{\'e}d{\'e}li, L.~Vivien, and M.~K. Smit, eds.), vol.~8767, p.~87670E, International Society for Optics and Photonics, SPIE, 2013.

\bibitem{beens_comparing_2019}
M.~Beens, M.~L.~M. Lalieu, A.~J.~M. Deenen, R.~A. Duine, and B.~Koopmans, ``Comparing all-optical switching in synthetic-ferrimagnetic multilayers and alloys,'' {\em Phys. Rev. B}, vol.~100, no.~22, p.~220409, 2019.

\bibitem{cao_dynamics_2020}
A.~Cao, Y.~L.~W. van Hees, R.~Lavrijsen, W.~Zhao, and B.~Koopmans, ``Dynamics of all-optically switched magnetic domains in {Co}/{Gd} heterostructures with {Dzyaloshinskii}-{Moriya} interaction,'' {\em Phys. Rev. B}, vol.~102, p.~104412, Sept. 2020.

\bibitem{bogart_dependence_2009}
L.~K. Bogart, D.~Atkinson, K.~O’Shea, D.~McGrouther, and S.~McVitie, ``Dependence of domain wall pinning potential landscapes on domain wall chirality and pinning site geometry in planar nanowires,'' {\em Phys. Rev. B}, vol.~79, p.~054414, Feb. 2009.

\bibitem{finocchio_promise_2021}
G.~Finocchio, M.~Di~Ventra, K.~Y. Camsari, K.~Everschor-Sitte, P.~Khalili~Amiri, and Z.~Zeng, ``The promise of spintronics for unconventional computing,'' {\em J. Magn. Magn. Mater.}, vol.~521, p.~167506, Mar. 2021.

\bibitem{incorvia_spintronics_2024}
J.~A.~C. Incorvia, T.~P. Xiao, N.~Zogbi, A.~Naeemi, C.~Adelmann, F.~Catthoor, M.~Tahoori, F.~Casanova, M.~Becherer, G.~Prenat, and S.~Couet, ``Spintronics for achieving system-level energy-efficient logic,'' {\em Nat. Rev. Electr. Eng.}, vol.~1, pp.~700--713, Nov. 2024.

\bibitem{grollier_neuromorphic_2020}
J.~Grollier, D.~Querlioz, K.~Y. Camsari, K.~Everschor-Sitte, S.~Fukami, and M.~D. Stiles, ``Neuromorphic spintronics,'' {\em Nat. Electron.}, vol.~3, pp.~360--370, July 2020.

\bibitem{marrows_neuromorphic_2024}
C.~H. Marrows, J.~Barker, T.~A. Moore, and T.~Moorsom, ``Neuromorphic computing with spintronics,'' {\em npj Spintronics}, vol.~2, p.~12, Apr. 2024.

\bibitem{gasse_recent_2019}
K.~Van~Gasse, S.~Uvin, V.~Moskalenko, S.~Latkowski, G.~Roelkens, E.~Bente, and B.~kuyken, ``Recent advances in the photonic integration of mode-locked laser diodes,'' {\em IEEE Photonics Technol. Lett.}, vol.~31, no.~23, pp.~1870--1873, 2019.

\end{thebibliography}
\end{document}